%% file: AR-eigen-R2.tex
\documentclass[12pt]{article}
\usepackage[utf8]{inputenc}
\usepackage{graphicx}
\usepackage{amsmath, amssymb,theorem,verbatim}
\usepackage{hyperref}
\usepackage{textcomp} 
\usepackage{setspace}
\usepackage{mathtools}
\usepackage[round]{natbib}
\usepackage{multirow} 
\usepackage{authblk}
\usepackage[usenames,dvipsnames]{xcolor}

\numberwithin{equation}{section}

\newtheorem{prop}{Proposition}[section] 
\newtheorem{theorem}{Theorem}[section]
\newtheorem{lemma}{Lemma}[section] 
\newtheorem{assumption}{Assumption}[section] 
\newtheorem{defin}{Definition}[section] 
\newtheorem{corollary}{Corollary}[section] 

\newtheorem{remark}{Remark}[section]

\newcommand{\var}{\mathrm{var}}

\newcommand{\Ex}{\mathbb{E}}
\newcommand{\diag}{\mathrm{diag}}
\newcommand{\yunder}{\underline{y}}

\newcommand{\Dcon}{\stackrel{\mathcal{D}}{\rightarrow}}
\newcommand{\Pcon}{\stackrel{P}{\rightarrow}}

\newcommand{\N}{\mathbb{N}}

\newcommand{\R}{\mathbb{R}}

\title{On the edge eigenvalues of the precision matrices of
nonstationary autoregressive processes}

\author{Junho Yang  \thanks{\textit{Email}: \url{junhoyang@stat.sinica.edu.tw}}
}
\affil{Institute of Statistical Science, Academia Sinica}

\date{\today}

\begin{document}

\maketitle 

\begin{abstract}
This paper investigates structural changes in the parameters of first-order autoregressive models by analyzing the edge eigenvalues of the precision matrices. Specifically, edge eigenvalues in the precision matrix are observed if and only if there is a structural change in the autoregressive coefficients. We show that these edge eigenvalues correspond to the zeros of a determinantal equation. Additionally, we propose a consistent estimator for detecting outliers within the panel time series framework, supported by numerical experiments.

\vspace{1em}

\end{abstract}

\section{Introduction and main results} \label{sec:intro}

Let $\{y_t: t \in \mathbb{N}\}$ be a mean-zero time series with the recursion:
\begin{equation} \label{basic_model}
y_t = \rho_t y_{t-1} + z_{t}, \quad t \in \mathbb{N},
\end{equation}
where the initial value is set to $y_0 = 0$ and $\{ z_t: t \in \mathbb{N} \}$ is a white noise process with unit variance, i.e., $\var(z_t) = 1$.
The above recursive model is known as the autoregressive (AR) model of order one, denoted as AR$(1)$.
In the special scenario where $\{\rho_{t}\}$ are constants with absolute values less than one, the process $\{y_t\}$ converges to a causal stationary AR process. Consequently, we refer to this model as a stationary AR$(1)$ process.

Given a realization of an AR$(1)$ process, our focus lies in the structure of its AR coefficients. To establish the asymptotic theory, for each $n \in \N$, we observe a time series $\underline{y}_n = (y_{1,n}, \dots, y_{n,n})^\top \in \R^{n}$ with the following recursive relationship:
\begin{equation} \label{eq:tri-AR}
y_{t,n} = \rho_{t,n} y_{t-1,n} + z_{t,n}, \quad t \in \{1, \dots, n\},
\end{equation}
where $y_{0,n} = 0$ is the initial value, and the AR coefficients are given by:
\begin{equation} \label{eq:rho.model}
\rho_{t,n} = \rho + \sum_{j=1}^{m} \varepsilon_j I_{E_{j,n}}(t), \quad t \in \{1, \dots, n\}.
\end{equation}
Here, $\rho \in (-1,1) \setminus \{0\}$ denotes the baseline AR coefficient, $m \in \mathbb{N} \cup \{0\}$ is the number of breakpoints, and for positive $m \in \mathbb{N}$, $\{\varepsilon_j\}_{j=1}^{m}$ represent nonzero perturbations, $\{E_{j,n}\}_{j=1}^{m}$ denote disjoint intervals on $\{1, \dots, n\}$, and $I_A(t)$ denotes the indicator function. In (\ref{eq:rho.model}), we adopt the convention $\sum_{j=1}^{0} \varepsilon_j I_{E_{j,n}}(t) = 0$. Consequently, when $m = 0$, we have a stationary AR$(1)$ model, referred to as the \textit{null model}. However, for $m > 0$, the associated AR$(1)$ process is no longer stationary due to structural changes in the AR coefficients. Consequently, we denote the case $m > 0$ as the \textit{Structural Change Model} (SCM) or the alternative model. We refer interested readers to \cite{p:ding-23} for a general class of time-varying structures of nonstationary time series and its global approximation by AR processes.

Statistical inference for $H_0: m = 0$ against $H_A: m > 0$ was initially developed by \cite{p:pag-55} using the cumulative sum (CUSUM) method. While the original CUSUM procedure was intended to detect changes in the mean structure of independent samples, its approach was soon extended to time series data (see, e.g., \cite{p:bag-77, p:dav-95, p:lee-03, p:gom-08, p:gom-ser-09}, and \cite{p:aue-13}). The test procedures outlined in the aforementioned literature rely on likelihood ratio or Kolmogorov-Smirnov type tests, necessitating the assumption
\begin{equation} \label{eq:trad}
\lim_{n \rightarrow \infty} \frac{|E_{j,n}|}{n} = \tau_j \in (0,1), \quad j \in \{1, \dots, m\},
\end{equation}
to distinguish the asymptotic behaviors of the test statistics under the null and alternative models.

However, in many real-world applications, particularly those involving economic data, changes often occur sporadically, meaning $|E_{j,n}| = o(n)$ as $n \rightarrow \infty$. In extreme cases, one may have $\sup_{n \in \mathbb{N}} |E_{j,n}| < \infty$ for all $j \in \{1, \dots, m\}$. To detect such abrupt changes, \cite{p:fox-72} considered outliers in Gaussian autoregressive moving average (ARMA) models and proposed a likelihood ratio test. See also \cite{p:hil-82, p:cha-88, p:tsa-88}. In these outlier models, the time series $\{Y_t\}$ takes the form
\begin{equation} \label{eq:disturbedARIMA}
Y_t = \omega_0 \frac{\omega(B)}{\delta(B)} e_t^{(d)} + Z_t, \quad t \in \mathbb{Z},
\end{equation}
where $\{Z_t: t \in \mathbb{Z}\}$ represents an unobserved Gaussian ARMA process, $\omega_0$ denotes the scale, $\omega(\cdot)$ and $\delta(\cdot)$ are polynomials, $B$ is the backshift operator, and $e_{t}^{(d)}$ is either a deterministic or stochastic component. However, there is no clear connection between the SCM under consideration and the outlier model in (\ref{eq:disturbedARIMA}).

\begin{figure}[h]
\centering
{{\includegraphics[page=1,width=0.48\textwidth]{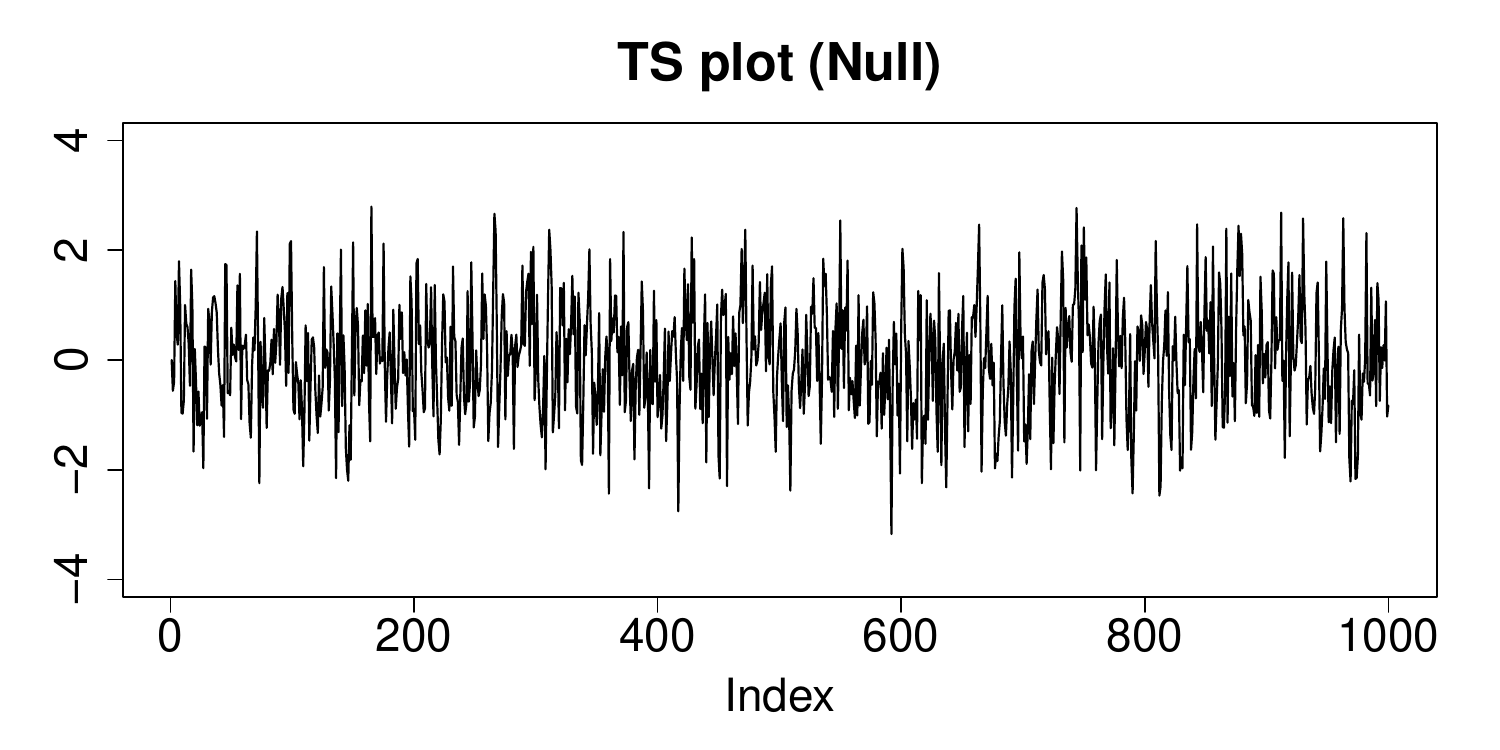} }}%
{{\includegraphics[page=2,width=0.48\textwidth]{tsplot2.pdf} }}
\caption{\textit{Time series trajectories for the null model (left) and the SCM (right). Vertical dashed line indicates the time of the structural change in the SCM.}}%
\label{fig:tsplot}%
\end{figure}

The aim of this article is to propose a new approach to characterize structural changes in parameters when $|E_{j,n}| = o(n)$ as $n \rightarrow \infty$. Consequently, within the framework under consideration, conventional likelihood-based approaches are not suitable. Instead, our focus shifts towards defining
\begin{equation} \label{eq:Andefition}
A_n = [\var(\underline{y}_n) ]^{-1} \in \R^{n \times n}, \quad n \in \N,
\end{equation}
which is the inverse covariance matrix (often referred to as the precision matrix) of $\underline{y}_n$. Given that $\var(\underline{y}_n)$ is symmetric and positive definite, so is $A_n$. Therefore, we can define the empirical spectral distribution (ESD) of $A_{n}$ as
\begin{equation}\label{eq:emfdis}
\mu_{A_{n}} := \frac{1}{n} \sum_{i=1}^{n} \delta_{\lambda_i(A_{n})}, \quad n\in \mathbb{N},
\end{equation}
where $\delta_{x}$ denotes the point mass at $x \in \mathbb{R}$, and $\lambda_{1} (A_n) \geq \ldots \geq \lambda_n (A_n) > 0$ denote the positive eigenvalues of $A_n$ arranged in decreasing order. To motivate the behavior of the ESD, we consider the following two scenarios regarding the AR coefficients:
\begin{equation} \label{eq:sim1}
\text{Null}: \rho_{t,1000} = 0.3 \quad \text{and} \quad
\text{SCM}: \rho_{t,1000} = 0.3 + 0.2 I_{\{50\}}(t), \quad t \in \{1, \dots, 1000\}.
\end{equation} 
Therefore, there is one and only one change that occurs at $t=50$ in SCM. Figure \ref{fig:tsplot} shows a realization of the time series under both the null model (left panel) and the SCM (right panel). A vertical line in the right panel denotes the occurrence of the structural change. It is apparent from these plots that it is hardly distinguishable to identify the presence of a structural change in SCM.

\begin{figure}[h]
\centering
{{\includegraphics[page=1,scale=0.48]{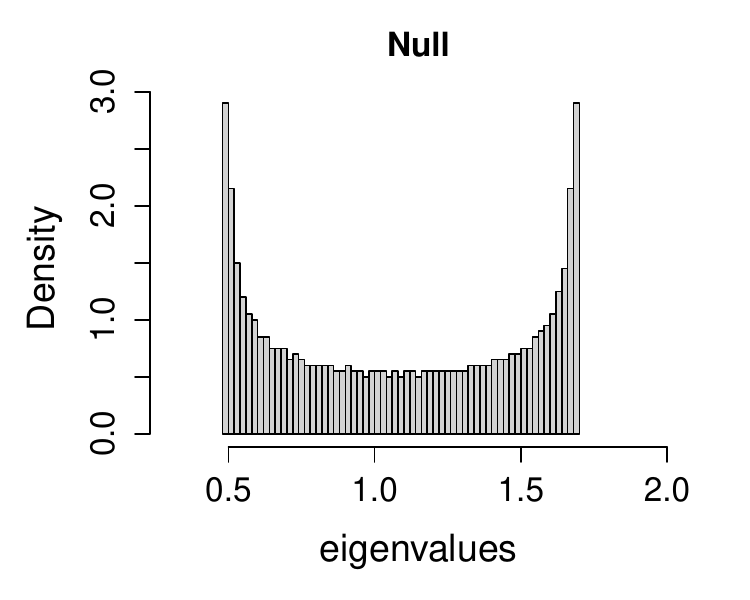} }}%
\quad
{{\includegraphics[page=2, scale=0.48]{eigendist.pdf} }}
\caption{\textit{Empirical spectral distributions the precision matrices. Left: null model, right: the SCM. Crosses on the right panel indicate the outliers.}}%
\label{fig:empdist}%
\end{figure}

In contrast, Figure \ref{fig:empdist} compares the ESDs of the null model (left panel) and the SCM (right panel). It is evident that:
\begin{itemize}
\item[(i)] The ESDs of the two models are almost identical.
\item[(ii)] Under the SCM, two outliers (indicated by crosses) are observed, which are separated from the bulk spectrum.
\end{itemize}
Hence, it becomes apparent that spectral statistics may offer an important means to characterize the behaviors of the SCM. 

To rigorously argue the first observation, we study the limiting spectral distribution (LSD) of $\mu_{A_n}$ under both the null model and SCM. Let $\mu_{\rho}$ be a probability measure on $\mathbb{R}$ that depends solely on $\rho \in (-1,1) \setminus \{0\}$ in (\ref{eq:rho.model}), with distribution
\begin{equation} \label{eq:cdf}
F_{\mu_\rho} (t) = \mu_\rho ((-\infty,t]) = \frac{1}{2\pi} \int_{0}^{2\pi} I_{(-\infty, t]}(1+\rho^2-2\rho \cos x) \, dx, \quad t \in \mathbb{R}.
\end{equation}
Our result, corresponding to the first observation, is as follows.
\begin{theorem} \label{thm:1}
Let $A_n$ be the precision matrix of an AR$(1)$ model, where the AR coefficients adhere to (\ref{eq:rho.model}) (including the case $m=0$). Furthermore, we assume
\begin{equation} \label{eq:Ejn-0}
\lim_{n \rightarrow \infty} |E_{j,n}|/n =0, \quad j \in \{1, \dots, m\}.
\end{equation}
Then, we have $\mu_{A_n} \xrightarrow{D} \mu_{\rho}$ as $n \rightarrow \infty$, where $\xrightarrow{D}$ denotes weak convergence. Moreover, the LSD $\mu_{\rho}$ is compactly supported.
\end{theorem}
The above theorem asserts that under the condition (\ref{eq:Ejn-0}), the ESDs of both the null model and SCM converge to the same distribution.

Now, we turn our attention to the second observation. Given a sequence of Hermitian matrices $\{X_n\}$, below we define the set of `outliers' of $\{X_n\}$. 
\begin{defin} \label{def:outlier}
Let $\{X_n\}$ be a sequence of Hermitian matrices and the ESDs $\{\mu_{X_n}\}$ converge weakly to some probability measure $\mu$ with a compact support $S_\mu \subset \mathbb{R}$. Then, $x \in \mathbb{R} \setminus S_{\mu}$ is called an outlier of the sequence $\{X_n\}$ (or, an outlier of the sequence $\{\mu_{X_n}\}$) if there exists $j \in \mathbb{N}$ such that
\begin{equation} \label{eq:limdist}
\lim_{n \to \infty} \lambda_{j}(X_n) = x \quad \text{or} \quad \lim_{n \to \infty} \lambda_{n+1-j}(X_n) = x.
\end{equation}
We denote $out (\{X_n\})$ as the set of all outliers of $\{X_n\}$ (or, outliers of $\{\mu_{X_n}\}$). 
\end{defin}
Heuristically, the outliers of $\{X_n\}$ represent the limiting edge eigenvalues of $X_n$ that lie outside the bulk spectrum. We also mention that the concept of outliers resembles that of the spiked model in the random matrix literature (refer to \cite{p:bai-06, p:ding-21} for a review). In our situation, the defined outliers have an $O(1)$ separation from the bulk spectrum.

Now, we study the outliers of the null model and SCM. The following theorem addresses the outlier behavior in the null model.
\begin{theorem} \label{lemma:outlier.null}
Let $\{A_{0,n}\}$ be a sequence of precision matrices corresponding to the null model. Then,
\begin{equation*}
out(\{ A_{0,n}\} ) = \emptyset.
\end{equation*}
\end{theorem}
Therefore, as anticipated in the left panel of Figure \ref{fig:empdist}, the ESD of the null model does not include any outliers.

Next, we focus on the outliers of the SCM. In Sections \ref{subsec:ssc}--\ref{subsec:block}, we prove that under mild conditions, we have
\begin{equation} \label{eq:outSCM}
out(\{A_{n}\}) \neq \emptyset.
\end{equation}
In particular, for the simplest case where $m = 1$ and $|E_{1,n}| = 1$ for all $n \in \mathbb{N}$, we derive the closed-form expressions for the two (and only two) outliers, as observed in the right panel of Figure \ref{fig:empdist}. Please refer to Theorem \ref{thm:spacing} for further details. In the general case, we show that $out(\{A_{n}\})$ can be fully determined by the zeros of a determinantal equation, making the numerical evaluation of the entries in $out(\{A_{n}\})$ feasible.

The remainder of the article is structured as follows. In Section \ref{sec:prelim}, we introduce the notation used throughout the article and present a preliminary result concerning the expression of $A_n$ in (\ref{eq:Andefition}). Section \ref{sec:LSD} investigates the LSD of both the null model and SCM, and proves Theorem \ref{thm:1}. In Section \ref{sec:single}, we define the outliers of $\{A_n\}$ and demonstrate their absence in the null model (Section \ref{subsec:out}). Sections \ref{subsec:ssc}--\ref{subsec:block} explore the outliers of the SCM, covering the single structural change model (Section \ref{subsec:ssc}), the single-interval structural change model (Section \ref{subsec:interval}), and the general structural change model (Section \ref{subsec:block}). Section \ref{sec:iden} addresses the identifiability of parameters in the SCM. In Section \ref{sec:est}, we propose a consistent estimator for $out(\{A_{n}\})$ within the panel time series framework and demonstrate its efficacy through numerical experiments. Section \ref{thm:spacing:proof} provides a proof for Theorem \ref{thm:spacing}, which derives a closed-form expression for the two outliers in the single structural change model. 

Lastly, additional properties of the LSD, supplementary proofs, and technical lemmas are presented in the Appendix.

\section{Limiting spectral distribution of $\mu_{A_n}$} \label{sec:asymp}

\subsection{Preliminaries} \label{sec:prelim}

In this section, we introduce the notation and terminology used in the article and provide a preliminary result. For the SCM, we denote the disjoint intervals $\{E_{j,n}\}_{j=1}^{m}$ as
\begin{equation*}
E_{j,n} = [k_{j,n}, k_{j,n} + h_{j,n} - 1] \cap \mathbb{N}, \quad j \in \{1, \dots, m\},
\end{equation*}
where we have an ordering
\begin{equation*}
1 \leq k_{1,n} < k_{1,n} + h_{1,n} - 1 < k_{2,n} < \dots < k_{m,n} < k_{m,n} + h_{m,n} - 1 \leq n.
\end{equation*}
Here, $m$ denotes the \textit{number of changes}, $k_j$ denotes the $j$th \textit{breakpoint}, $h_j$ denotes the $j$th \textit{length of change}, and $\varepsilon_j$ denotes the $j$th \textit{magnitude of change}. In particular, when $m=1$ and $h_{1,n} = 1$, we omit the subscript "1" in $k_{1,n}$ and $\varepsilon_{1}$, and express
\begin{equation} \label{eq:sscm}
\rho_{t,n} = \rho + \varepsilon I_{k_{n}} (t), \quad t \in \{1, \dots, n\}.
\end{equation}
We refer to (\ref{eq:sscm}) as the single structural change model (single SCM).

Let $A_n$ be a general precision matrix of an AR$(1)$ process. Sometimes, it is necessary to differentiate between the null model ($m=0$) and the SCM ($m>0$). In such cases, we denote $A_{0,n}$ and $B_n$ as the precision matrices for the null model and the SCM, respectively. For a real symmetric matrix $A \in \mathbb{R}^{n \times n}$, $spec(A) = \{\lambda_i(A)\}_{i=1}^{n}$ denotes the spectrum of $A$. Lastly, $\wedge$ and $\vee$ denote the minimum and maximum operations, respectively.

The following lemma provides an explicit form of the entries of $A_n$.
\begin{lemma}
\label{prop:explicit}
Let $A_n$ be a precision matrix of AR$(1)$ model. Then, $A_n$ is an $n \times n$ symmetric tri-diagonal matrices with entries
\begin{equation} \label{eq:Anform}
[A_n]_{i,j} = \left \{\begin{array}{ll}
1, & i = j = n.\\
1+ \rho_{i+1,n}^2, & i = j \in \{1, \dots, n-1\}.\\
- \rho_{i \vee j,n}, & |i-j| = 1. \\
0, & o.w. \end{array}
\right. 
\end{equation}
\end{lemma}
\noindent \textit{Proof}. 
Let $\underline{z}_n = (z_{1,n}, \dots, z_{n,n})^\top$, where $\{z_{t,n}\}$ are as defined in (\ref{eq:tri-AR}). Then, $\var(\underline{z}_n) = I_n$, the identity matrix of order $n$. Using the recursive formula in (\ref{eq:tri-AR}), we obtain the following linear equation:
\begin{equation*}
\underline{z}_n = L_n \yunder_n,
\quad \text{where} \quad
L_n = \begin{pmatrix}
1 & 0 & \cdots & 0\\
-\rho_{2,n} & 1 & \cdots & 0\\
\vdots & \ddots & \ddots & 0\\
0 & 0 & -\rho_{n,n} & 1
\end{pmatrix} \in \mathbb{R}^{n \times n}. 
\end{equation*}
Taking the variance on each side of the above equation and using the fact that
$A_n = L_n^{\top} [\var(\underline{z}_n)]^{-1} L_n$, we obtain (\ref{eq:Anform}).
\hfill $\Box$

\subsection{LSDs of the null model and SCM and proof of Theorem \ref{thm:1}} \label{sec:LSD}
Let $A_{0,n}$ be the precision matrix of the null model with the constant AR coefficient $\rho \in (-1,1) \setminus \{0\}$. Then, according to Lemma \ref{prop:explicit}, $A_{0,n}$ can be explicitly written as
\begin{equation} \label{eq:An.null}
A_{0,n} = \begin{pmatrix}
1 + \rho^2 & -\rho & 0 & \cdots & 0\\
-\rho & 1 + \rho^2 & -\rho & \ddots & \vdots \\
0 & -\rho & \ddots & \ddots & 0\\
\vdots & \ddots & \ddots & 1 + \rho^2 & -\rho \\
0 & \cdots & 0 & -\rho & 1
\end{pmatrix} \in \mathbb{R}^{n \times n}, \quad n \in \mathbb{N}.
\end{equation}

For technical reasons, we define the slightly perturbed matrix
\begin{equation}
\label{eq:perturb}
\widetilde{A}_{0,n} = A_{0,n} + \rho^2 E_n, \quad n \in \mathbb{N},
\end{equation}
where $E_n = \diag(0, \dots, 0, 1) \in \mathbb{R}^{n \times n}$. Since $\widetilde{A}_{0,n}$ is a tri-diagonal Toeplitz matrix, one can derive explicit expressions for the entire set of eigenvalues and the corresponding eigenvectors of $\widetilde{A}_{0,n}$ (cf. \cite{p:str-83}, Proposition 2). Specifically, for $\rho \in (0,1)$, the $k$th smallest eigenvalue is given by
\begin{equation}
\label{eq:evalperturb}
\lambda_{n+1-k}(\widetilde{A}_{0,n}) = 1 - 2\rho \cos\left( \frac{k \pi}{n+1} \right) + \rho^2, \quad k \in \{1, \dots, n\},
\end{equation}
and the corresponding normalized eigenvector is $\widetilde{u}_{n+1-k} = (\widetilde{u}_{1,n+1-k}, \dots, \widetilde{u}_{n,n+1-k})^{\top}$, where
\begin{equation}
\label{eq:evecperturb}
\widetilde{u}_{j,n+1-k} = \sqrt{\frac{2}{n+1}} \sin\left( \frac{kj \pi}{n+1} \right), \quad j \in \{1, \dots, n\}.
\end{equation}
In the case when $\rho \in (-1,0)$, the eigenstructure has a similar expression but is arranged in reverse order. Now, since $\widetilde{A}_{0,n}$ is a Toeplitz matrix, we can directly apply the Szeg{\"o} limit theorem to $\widetilde{A}_{0,n}$ (cf. \cite{p:gre-58}, Chapter 5).

\begin{lemma} \label{lemma:LSD}
Let $\widetilde{A}_{0,n}$ be defined as in (\ref{eq:perturb}). Let $\mu_{\rho}$ be the probability measure on $\mathbb{R}$ with distribution (\ref{eq:cdf}).
Then,
\begin{equation} \label{eq:LSD-null}
\mu_{\widetilde{A}_{0,n}} \Dcon \mu_{\rho}, \quad n \rightarrow \infty.
\end{equation}
Furthermore, $\mu_{\rho}$ has support of the form $[a_\rho, b_\rho] \subset \mathbb{R}$, where
\begin{equation} \label{eq:bounds}
a_{\rho} = \inf(\text{supp}(\mu_{\rho})) = (1 - |\rho|)^2 \quad \text{and} \quad
b_{\rho} = \sup(\text{supp}(\mu_{\rho})) = (1 + |\rho|)^2.
\end{equation}
\end{lemma}
\noindent \textit{Proof}. 
Since the eigenvalues of $\widetilde{A}_{0,n}$ are provided in (\ref{eq:evalperturb}), (\ref{eq:LSD-null}) follows immediately from the Szeg{\"o} limit theorem together with (\ref{eq:cdf}). (\ref{eq:bounds}) is also clear since the range of $(1 + \rho^2 - 2\rho \cos x)$ in (\ref{eq:cdf}) is $[(1 - |\rho|)^2, (1 + |\rho|)^2]$.
\hfill $\square$

From the lemma above, we observe that a slightly perturbed matrix of the null model has an LSD $\mu_{\rho}$. Next, we explore the LSDs (if they exist) of the null model and SCM. Note that the null model is a special case of (\ref{eq:rho.model}) by setting $m=0$; thus, it suffices to examine the LSD of the general SCM based on (\ref{eq:rho.model}).

The following theorem is key to the proof of Theorem \ref{thm:1}.
\begin{theorem} \label{prop:asymdist}
Let $A_n$ be a precision matrix of an AR$(1)$ model where the AR coefficients satisfy (\ref{eq:rho.model}), allowing for the case $m=0$. Let $\lim_{n \rightarrow \infty} |E_{j,n}|/n = \tau_j \in [0,1]$, $j \in \{1, \ldots, m\}$. Furthermore, we assume $|\rho + \varepsilon_j| < 1$, $j \in \{1, \ldots, m\}$. Then,
\begin{equation} \label{eq:LSD}
\mu_{A_n} \Dcon \big( 1 - \sum_{j=1}^{m} \tau_j \big) \mu_{\rho} + \sum_{j=1}^{m} \tau_j \mu_{\rho + \varepsilon_j}, \quad n \rightarrow \infty.
\end{equation}
\end{theorem}
\noindent \textit{Proof}. See Appendix \ref{prop:asymdist:proof}.
\hfill $\Box$

\vspace{0.5em}

Using the above theorem, we now can prove Theorem \ref{thm:1}.

\vspace{0.5em}

\noindent \textit{Proof of Theorem \ref{thm:1}}. 
Since $\tau_j = 0$ for $j \in \{1, \dots, m\}$ due to (\ref{eq:Ejn-0}), the assertion immediately follows from (\ref{eq:LSD}). Moreover, the compact support of $\mu_{\rho}$ is directly available from Lemma \ref{prop:asymdist}.
\hfill $\Box$

\section{Outliers of the SCM} \label{sec:single}
In this section, we study the outliers of the null model and SCM. Throughout the section, we assume that $|E_{j,n}| = h_{j,n} = h_j$ is fixed for all $n \in \mathbb{N}$, thus immediately satisfying (\ref{eq:Ejn-0}). Recall Definition \ref{def:outlier}. By Theorem \ref{thm:1}, the LSDs of $\{A_{0,n}\}$ (corresponding to the null model) and $\{B_n\}$ (corresponding to the SCM) are the same and equal to $\mu_\rho$. Moreover, since the support of $\mu_{\rho}$ is $S_{\rho} = [a_\rho, b_\rho]$, where $a_{\rho}$ and $b_{\rho}$  are given in (\ref{eq:bounds}), we can define the set of left and right outliers respectively by
\begin{equation*}
out_{L} (\{A_n\}) = out (\{A_n\}) \cup (-\infty, a_\rho) \quad \text{and} \quad
out_{R} (\{A_n\}) = out (\{A_n\}) \cap (b_\rho, \infty).
\end{equation*}

\subsection{Outlier behavior of the null model and proof of Theorem \ref{lemma:outlier.null}} \label{subsec:out}
In this section, we study the outliers of the null model. The following lemma states the behavior of the edge eigenvalues of $A_{0,n}$.
\begin{lemma} \label{lemma:tight}
Let $A_{0,n}$ be as defined in (\ref{eq:An.null}) and let $a_\rho$ and $b_\rho$ be as defined in (\ref{eq:bounds}).
Then, for fixed $j \in \mathbb{N}$, 
\begin{equation*}
\lim_{n \to \infty} \lambda_{n+1-j}(A_{0,n}) = a_\rho \quad \text{and} \quad
\lim_{n \to \infty} \lambda_{j}(A_{0,n}) = b_\rho.
\end{equation*}
\end{lemma}
\noindent \textit{Proof}. 
We will only prove the lemma for $\rho \in (0,1)$ and the $j$th smallest eigenvalue $\lambda_{n+1-j}(A_{0,n})$. The cases when $\rho \in (-1,0)$ and $\lambda_{j}(A_{0,n})$ can be treated similarly. Let 
\[
\alpha_{jn} = \frac{\lambda_{n+1-j}(A_{0,n}) - 1 - \rho^2}{2\rho}.
\]
Then, $\alpha_{jn} \in \big(\cos \frac{(n-j+1)\pi}{n}, \cos \frac{(n-j+1)\pi}{n+1}\big)$ due to \cite{p:str-83}, Proposition 1.
Therefore, for fixed $j \in \mathbb{N}$, $\lim_{n \rightarrow \infty} \alpha_{jn} = -1$, and in turn, $\lim_{n \rightarrow \infty} \lambda_{n+1-j}(A_{0,n}) = 1 - 2\rho + \rho^2 = a_\rho$. Thus, we obtain the desired result. \hfill $\Box$

\vspace{0.5em}

The above lemma shows that under the null model, the $j$th smallest (resp. largest) eigenvalue of $A_{0,n}$ converges to the lower (resp. upper) bound of the support of the LSD of $A_{0,n}$. Consequently, we can prove Theorem \ref{lemma:outlier.null}, which states that there are no outliers in the null model.

\vspace{0.5em}

\noindent \textit{Proof of Theorem \ref{lemma:outlier.null}}. 
The assertion is immediately followed by Lemma \ref{lemma:tight} and the definition of the outliers.
\hfill $\Box$

\subsection{Outliers of the single SCM} \label{subsec:ssc}

In this section, we investigate the outliers of the single SCM given by (\ref{eq:sscm}). To obtain the explicit form of the outliers, we require the following condition on the breakpoint:
\begin{equation} \label{assum:SCM}
\text{The breakpoint } k=k_n \text{ of the single SCM is such that } \lim_{n\to\infty} k_n = \lim_{n\to\infty} (n-k_n) = \infty.
\end{equation}
Next, let
\begin{equation}
\begin{aligned}
s &= \frac{ \rho \varepsilon (\varepsilon+2\rho) - \sqrt{ \rho^2 \varepsilon^2 (\varepsilon+2\rho)^2 + 4\rho^2 (\varepsilon +\rho)^2 }}{2 (\varepsilon + \rho)^2} \\
\text{and} \quad
t &= \frac{ \rho \varepsilon (\varepsilon+2\rho) + \sqrt{ \rho^2 \varepsilon^2 (\varepsilon+2\rho)^2 + 4\rho^2 (\varepsilon +\rho)^2 }}{2 (\varepsilon + \rho)^2}
\end{aligned}
\label{eq:st}
\end{equation}
be the roots of the quadratic equation
\begin{equation} \label{eq:sscm-quadratic}
(\varepsilon+\rho)^2 x^2 - \rho\varepsilon (\varepsilon+2\rho) x - \rho^2 = 0.
\end{equation}
The following theorem provides the closed-form solution for $out(\{B_{n}\})$ under the single SCM.

\begin{theorem}
\label{thm:spacing}
Let $\{B_n\}$ be the precision matrix of the single SCM. 
Furthermore, assume the breakpoint $k_n$ satisfies (\ref{assum:SCM}) and $\rho (\rho +\varepsilon) > 0$. Then, the following two assertions hold:
\begin{itemize}
\item If $|\rho| \geq |\rho+\varepsilon|$, then $out_{L}(\{ B_n\}) = \emptyset$.
\item If $|\rho| < |\rho + \varepsilon|$, then $out_{L}(\{ B_n\}) = \{\mathfrak{m}\}$ and $out_{R}(\{ B_n\}) = \{\mathfrak{M}\}$, where
\begin{equation} \label{eq:singleform1}
(\mathfrak{m}, \mathfrak{M}) = 
\begin{cases}
\left(1+ \rho^2 - \rho (s + s^{-1}), 1+ \rho^2 - \rho (t + t^{-1})\right), & \text{if } \rho \in (-1, 0). \\
\left(1+ \rho^2 - \rho (t + t^{-1}), 1+ \rho^2 - \rho (s + s^{-1})\right), & \text{if } \rho \in (0,1).
\end{cases}
\end{equation}
\end{itemize}
\end{theorem}
\noindent \textit{Proof}. See Section \ref{thm:spacing:proof}.
\hfill $\Box$

\begin{remark} \label{rmk:spacing}
\begin{enumerate}
\item[(i)] 
The dichotomy in Theorem \ref{thm:spacing} shows that if the modulus of the AR coefficient at the breakpoint is smaller than the baseline AR coefficient, then we cannot detect the outlier. Specifically, when $|\rho+\varepsilon| > |\rho|$, we observe exactly two outliers: one on the left and another on the right.

\item[(ii)] 
Suppose the breakpoint $k_n=k \in \mathbb{N}$ is fixed so that the condition (\ref{assum:SCM}) is not satisfied. Furthermore, assume $\rho (\rho + \varepsilon) > 0$ and $|\rho| < |\rho + \varepsilon|$. Then, there exists $c \in (0,1)$, depending only on $\rho$, such that
\begin{equation*}
|out_{L}(\{ B_n\}) - \mathfrak{m}| \leq c^{k} \quad \text{and} \quad
|out_{R}(\{ B_n\}) - \mathfrak{M}| \leq c^{k},
\end{equation*}
where $\mathfrak{m}$ and $\mathfrak{M}$ are given in (\ref{eq:singleform1}). The proof of this result can be found in Section \ref{thm:spacing:proof}, [step 7]. In practice, $k \geq 5$ is sufficiently large to approximate $out(\{ B_n\}) \approx \{\mathfrak{m}, \mathfrak{M}\}$.
\end{enumerate}
\end{remark}
\subsection{Outliers of the single interval SCM} \label{subsec:interval}

As an intermediate step, we consider a slight generalization of the single SCM, where the AR coefficients are given by
\begin{equation} \label{eq:interval}
\rho_{t,n} = \rho + \varepsilon I_{[k_n, k_n + h - 1]}(t), \quad t \in \{1, \dots, n\}.
\end{equation}
This represents a model with a single structural change occurring over a fixed interval length of $h \in \mathbb{N}$. We refer to this model as the single `interval' SCM. Note that for $h = 1$, the single interval SCM reduces to the single SCM.

We now investigate the outliers of the single interval SCM. To do so, we define several functions and values. First, let
\begin{equation} \label{eq:ftrans}
f(z) = 1 + \rho^2 - \rho \left(z + z^{-1}\right), \quad z \in (-1,1) \setminus \{0\}.
\end{equation}
It is straightforward that $f$ is a bijective mapping from $(-1,1) \setminus \{0\}$ to $[a_{\rho}, b_{\rho}]^{c}$, where $a_{\rho}$ and $b_{\rho}$ are defined in (\ref{eq:bounds}). Let $f^{-1}: [a_{\rho}, b_{\rho}]^c \rightarrow (-1,1) \setminus \{0\}$ denotes the inverse mapping of $f$.
Next, we consider the functions $\alpha(x)$, $\beta(x)$, and $\gamma(x)$, defined as follows:
\begin{equation} \label{eq:param}
\alpha(x) = \frac{ \rho x^{-1} + \varepsilon (\varepsilon + 2\rho)}{\varepsilon + \rho}, \quad
\beta(x) = \frac{\rho \left( x + x^{-1} \right) + \varepsilon (\varepsilon + 2\rho)}{\varepsilon + \rho}, \quad
\gamma(x) = \frac{\rho x^{-1}}{\varepsilon + \rho}.
\end{equation}

By using the aforementioned definitions, we introduce the sequence of tri-diagonal matrix functions $M_{h+1}(z)$, $h \in \mathbb{N}$, on $[a_{\rho},b_{\rho}]^{c}$ by
\begin{equation} \label{eq:M2}
M_2(z) = \begin{pmatrix}
\alpha(f^{-1}(z)) & -1 \\
-1 & \gamma(f^{-1}(z))
\end{pmatrix}, 
\end{equation} 
and for $h \in \{2, 3, \dots\}$,
\begin{equation} \label{eq:determ_multi}
M_{h+1}(z) = \begin{pmatrix}
\alpha(f^{-1}(z)) & -1 & & \\
-1 & \beta(f^{-1}(z)) & \ddots & & \\
& \ddots & \ddots & \ddots \\
& & \ddots & \beta(f^{-1}(z)) & -1 \\
& & & -1 & \gamma(f^{-1}(z))
\end{pmatrix} \in \mathbb{R}^{(h+1)\times(h+1)}.
\end{equation}
Below, we prove a connection between the outliers $(\mathfrak{m}, \mathfrak{M})$ of the single SCM (corresponding to $h=1$) and $M_2$.
\begin{prop} \label{prop:det2}
Let $\mathfrak{m}, \mathfrak{M}$ be defined as in (\ref{eq:singleform1}). Then, $\mathfrak{m}$ and $\mathfrak{M}$ are the solutions of the determinantal equation $\det M_2(z) = 0$. 
\end{prop}
\noindent \textit{Proof}. 
By simple algebra, it is easily seen that the solutions to $\det M_2(f(x)) = \alpha(x) \gamma(x) - 1 = 0$ coincide with those of the quadratic equation in (\ref{eq:sscm-quadratic}). By definition, the zeros of (\ref{eq:sscm-quadratic}) are $s$ and $t$ in (\ref{eq:st}), implying that the roots of $\det M_2(z) = 0$ are $f(s)$ and $f(t)$. Therefore, the proposition holds for the specified $\mathfrak{m}$ and $\mathfrak{M}$ in (\ref{eq:singleform1}).
\hfill $\Box$

\vspace{0.5em}

The above proposition asserts that the two outliers of the single SCM indeed coincide with the solutions of the determinantal equation $\det M_2(z) = 0$. Below, we show that this analogous relationship also holds for the outliers of the single interval SCM.
\begin{theorem}
\label{thm:multi_determinant}
Let $B_{n}$ be the precision matrix of the single interval SCM with change length $h \in \mathbb{N}$. Moreover, we assume that the breakpoint $k_n$ satisfies (\ref{assum:SCM}). Then, the following two statements are equivalent:
\begin{itemize}
\item[(i)] $z \in \text{out}(\{ B_n \})$.
\item[(ii)] $\det M_{h+1}(z) = 0$ for some $z \notin [a_\rho, b_\rho]$.
\end{itemize}
\end{theorem}
\noindent \textit{Proof}. See Appendix \ref{thm:multi_determinant:proof}. \hfill $\Box$




\subsection{Outliers of the general SCM} \label{subsec:block}

We now turn our attention to the outliers of the general SCM described by (\ref{eq:rho.model}). Recall the ordered disjoint intervals $\{ E_{j,n} := [k_{j,n}, k_{j,n}+h_{j-1,n}]\}_{j=1}^{m}$, where $k_{1} < \dots < k_{m}$. Furthermore, we assume that the length of change $h_{j,n}=h_j \in \mathbb{N}$ is fixed for all $n \in \mathbb{N}$.

To investigate the outliers of the general SCM, we introduce the concept of submodels. For each $j \in \{1, \dots, m\}$, we define $B_n^{(j)}$ as the precision matrix of the single interval SCM, with the corresponding AR coefficients $\rho_{t,n}^{(j)}$ given by
\begin{equation} \label{eq:subj}
\rho_{t,n}^{(j)} = \rho + \varepsilon_j I_{[k_{j,n}, k_{j,n}+h_j-1]}(t), \quad t \in \{1, \dots, n\}.
\end{equation}
Thanks to Theorem \ref{thm:multi_determinant}, $out(\{ B_n^{(j)}\})$ can be fully determined by the solution to a determinantal equation. For the general SCM, it is anticipated that the outliers comprise the union of outliers of the submodels. To substantiate this, we require the following assumption on the spacing of breakpoints.
\begin{assumption} \label{assum:SCM2}
For $j \in \{1, \dots, m+1\}$, let $\Delta_{j,n} = k_{j,n} - k_{j-1,n}$, where we set $k_{0,n}=0$ and $k_{m+1,n} = n$. Then, we have
\begin{equation} \label{eq:delta}
\lim_{n \to \infty} \Delta_{n}  = \lim_{n \to \infty} \min_{1 \leq j \leq m+1} \Delta_{j,n} = \infty.
\end{equation}
It is worth noting that when $m=1$, Assumption \ref{assum:SCM2} is equivalent to condition (\ref{assum:SCM}).
\end{assumption} 
The following theorem addresses the outliers of the general SCM model.
\begin{theorem} \label{thm:block}
Let $B_{n}$ be the precision matrix of the SCM as in (\ref{eq:rho.model}), where there are $m \in \mathbb{N}$ changes.
Moreover, assume Assumption \ref{assum:SCM2} holds. Then,
\begin{equation} \label{eq:blockout}
out(\{B_n\}) = \bigcup_{j=1}^{m} out(\{B_n^{(j)}\}).
\end{equation} 
\end{theorem}
\noindent \textit{Proof}. See Appendix \ref{thm:block:proof}. \hfill $\Box$

\vspace{0.5em}

The following corollary is a direct consequence of Theorems \ref{thm:spacing} and \ref{thm:block} (we omit the proof).

\begin{corollary} \label{coro:multipleSSCM}
Suppose the same set of assumptions as in Theorem \ref{thm:block} hold. Furthermore, for $j \in \{1, \dots, m\}$, assume $h_j = 1$, $\rho(\varepsilon_j + \rho) > 0$, and $|\rho + \varepsilon_j| > |\rho|$. Then, we have
\begin{equation*}
out_L(\{B_n\}) = \{ \mathfrak{m}_1, \dots, \mathfrak{m}_m \} \quad \text{and} \quad
out_R(\{B_n\}) = \{ \mathfrak{M}_1, \dots, \mathfrak{M}_m \},
\end{equation*}
where, for $j \in \{1, \dots, m\}$, the outlier pair $(\mathfrak{m}_j, \mathfrak{M}_j)$ are as in Theorem \ref{thm:spacing}, but with $\varepsilon$ replaced by $\varepsilon_j$.
\end{corollary}

\section{Parameter identification} \label{sec:iden}

Let $\rho \in (-1,1) \setminus \{0\}$, $m \in \mathbb{N}$, $\underline{\varepsilon} = (\varepsilon_{1}, \dots, \varepsilon_{m})$, $\underline{k}_n = (k_{1,n}, \dots, k_{m,n})$, and $\underline{h} = (h_{1}, \dots, h_{m})$ be a set of parameters for the SCM. In this section, we focus on the identifiability of $out(\{B_{n}\})$ given these parameters. Given that the baseline AR coefficient $\rho$ can be readily estimated using classical methods such as the Yule-Walker estimator, and considering that the outliers are independent of the breakpoints $\underline{k}_n$ due to Theorem \ref{thm:block}, we confine our parameter of interest to $\theta = (m, \underline{\varepsilon}, \underline{h})$.

Let $\sigma \in S_{m}$ be a permutation on $\{1, \dots, m\}$, and let $\underline{\varepsilon}_{\sigma} = (\varepsilon_{\sigma(1)}, \dots, \varepsilon_{\sigma(m)})$. $\underline{h}_{\sigma}$ is defined similarly. Then, it is easily seen from Theorem \ref{thm:block} that
\begin{equation*}
out(\{B_{n}\} \mid (m, \underline{\varepsilon}, \underline{h})) = out(\{B_{n}\} \mid (m, \underline{\varepsilon}_{\sigma}, \underline{h}_{\sigma})), \quad \sigma \in S_{m},
\end{equation*}
where $out(\{B_{n}\} \mid (m, \underline{\varepsilon}, \underline{h}))$ denotes the outliers of the SCM given the parameters $m$, $\underline{\varepsilon}$, and $\underline{h}$ (assuming $\rho$ is known and $\underline{k}_n$ satisfies Assumption \ref{assum:SCM2}).

Therefore, the set of outliers remains unchanged under permutations of $(\underline{\varepsilon}, \underline{h})$. Below, we show that in the particular scenario where $\underline{h} = (1, \dots, 1)$, the parameters $m$ and $\underline{\varepsilon}$ are identifiable up to permutations.

\begin{theorem} \label{prop:ident}
Suppose the same set of assumptions as in Theorem \ref{thm:block} hold. Moreover, assume that the length of changes is equal to one for all $j \in \{1, \dots, m\}$. Define $\mathcal{E}_\rho = (0, \infty)$ for $\rho > 0$ and $\mathcal{E}_\rho = (-\infty, 0)$ for $\rho < 0$. Now, suppose there exist $(m_i, \underline{\varepsilon}_i) \in \{0, 1, \dots\} \times \mathcal{E}_\rho^{m_i}$ for $i \in \{1, 2\}$, such that
\begin{equation*}
out(\{B_n\} \mid (m_1, \underline{\varepsilon}_1, \underline{h})) = out(\{B_n\} \mid (m_2, \underline{\varepsilon}_2, \underline{h})),
\end{equation*}
then we have $m_1 = m_2$ and $\underline{\varepsilon}_2 = (\underline{\varepsilon}_1)_{\sigma}$
for some permutation $\sigma \in S_{m_1}$.
\end{theorem}
\noindent \textit{Proof}. 
By Corollary \ref{coro:multipleSSCM}, $out(\{B_n\} \mid (m_1, \underline{\varepsilon}_1, \underline{h})) = out(\{B_n\} \mid (m_2, \underline{\varepsilon}_2, \underline{h}))$ implies $m_1 = m_2 = m$. 

Let $out_{L}(\{B_n\} \mid (m, \underline{\varepsilon}_1, \underline{h})) = out_{L}(\{B_n\} \mid (m, \underline{\varepsilon}_2, \underline{h})) = \{\mathfrak{m}_1, \dots, \mathfrak{m}_{m}\}$ and let $out_{R}(\{B_n\} \mid (m, \underline{\varepsilon}_1, \underline{h})) = out_{R}(\{B_n\} \mid (m, \underline{\varepsilon}_2, \underline{h})) = \{\mathfrak{M}_1, \dots, \mathfrak{M}_{m}\}$, where $0 < \mathfrak{m}_1 \leq \dots \leq \mathfrak{M}_m < a_\rho$ and $b_\rho < \mathfrak{M}_{m} \leq \dots \leq \mathfrak{M}_1$. Then, by Proposition \ref{prop:det2}, there exists a permutation $(j_1, \dots, j_m) \in S_m$, such that for each $i \in \{1, \dots, m\}$, there exists $\varepsilon_i \in \mathcal{E}_\rho$ such that $f^{-1}(\mathfrak{m}_i)$ and $f^{-1}(\mathfrak{M}_{j_i})$ are the zeros of the quadratic equation
\begin{equation} \label{eq:quad}
-(\varepsilon_i + \rho)^2 z^2 + \varepsilon_i \rho (\varepsilon_i + 2\rho) z + \rho^2 = 0.
\end{equation}
We denote $\mathfrak{m}_i = \mathfrak{m}(\varepsilon_i)$ and $\mathfrak{M}_{j_i} = \mathfrak{M}(\varepsilon_i)$ to indicate the dependence on $\varepsilon_i$. After some algebra, for $\rho > 0$, it can be shown that $\mathfrak{m}(\varepsilon)$ is a decreasing function and $\mathfrak{M}(\varepsilon)$ is an increasing function of $\varepsilon \in \mathcal{E}_\rho$. Hence, if $\mathfrak{m}_1 \leq \dots \leq \mathfrak{M}_m$, then $\varepsilon_1 \geq \dots \geq \varepsilon_{m}$, leading to $\mathfrak{M}_{j_1} \geq \dots \geq \mathfrak{M}_{j_{m}}$. Consequently, we have $(j_1, \dots, j_m) = (1, 2, \dots, m)$. Similarly, for $\rho < 0$, it can also be shown that $(j_1, \dots, j_m) = (1, \dots, m)$.

Lastly, given an outlier pair $(\mathfrak{m}_i, \mathfrak{M}_i)$, the magnitude of change $\varepsilon_i \in \mathcal{E}_\rho$ is uniquely determined by (\ref{eq:quad}). Consequently, the sets $\underline{\varepsilon}_1$ and $\underline{\varepsilon}_2$ are identical up to some permutation. Thus, we obtain the desired result.
\hfill $\Box$

\begin{remark}[Conjecture on the General Case]
Although we do not yet have a proof, we conjecture that the results of Theorem \ref{prop:ident} are also true for the general length of changes $\underline{h}$. Specifically, let $\theta = (m, \underline{\varepsilon}, \underline{h})$, and for $\sigma \in S_m$, let $\theta_\sigma = (m, \underline{\varepsilon}_\sigma, \underline{h}_\sigma)$. Suppose there exist $\theta_1$ and $\theta_2$ such that $out(\{B_n\} \mid \theta_1) = out(\{B_n\} \mid \theta_2)$. We conjecture that $m_1 = m_2$ and there exists a permutation $\sigma \in S_{m_1}$ such that $\theta_2 = \theta_1^\sigma$.
\end{remark}

\section{Outlier detection of a panel time series} \label{sec:est}
In this section, we leverage the theoretical results established in Section \ref{sec:single} within the context of a panel time series. Consider the panel autoregressive model given by:
\begin{equation}
\label{eq:panel}
y_{j,t,n} = \rho_{t,n} y_{j,t-1,n} + z_{j,t,n},  \quad n \in \mathbb{N}, \quad t \in \{1, \dots, m\}, \quad j \in \{1, \dots, B\}.
\end{equation}
Here, we assume $y_{j,0,n}=0$, $\{ z_{j,t,n} \}$ are \textit{i.i.d.} random variables with mean zero and variance one, and $\{\rho_{t,n}\}$ are the common AR coefficients across $j$ that follow the SCM described in (\ref{eq:rho.model}).

Let $\underline{y}_{j;n} = (y_{j,1,n}, \dots, y_{j,n,n})^{\top} \in \mathbb{R}^{n}$ be the $j$th observed time series, with a common variance matrix $\text{var}(\underline{y}_{j;n}) = \Sigma_{n} \in \mathbb{R}^{n \times n}$. Our objective is to construct a consistent estimator for $out(\{\Omega_{n}\})$, where $\Omega_{n} = (\Sigma_{n})^{-1}$.
To achieve this, we first investigate a consistent estimator for $\Omega_{n}$. A natural plug-in estimator for $\Sigma_{n}$ is given by
\begin{equation} 
\label{eq:Sigmanb}
\widehat{\Sigma}_{n,B} = B^{-1} \sum_{j=1}^{B} (\underline{y}_{j;n} - \overline{y}_{j,n} \mathbf{1}_n) (\underline{y}_{j;n} - \overline{y}_{j,n} \mathbf{1}_n)^{\top} \in \mathbb{R}^{n \times n},
\end{equation}
where $\overline{y}_{j,n} = \frac{1}{n}\sum_{t=1}^{n} y_{j,t,n}$ and $\mathbf{1}_n$ is a column vector of ones.

However, as stated in \cite{p:wu-pou-09}, when $n$ increases at the same rate as $B$, the estimator $\widehat{\Sigma}_{n,B}$ no longer consistently estimates $\Sigma_{n}$. To obtain consistent estimation of $\Sigma_n$ and $\Omega_n$ in the high-dimensional regime (where $n \gg B$), one typically needs to assume some sparsity condition on $\Omega_n$, which, fortunately, holds in our framework due to Lemma \ref{prop:explicit}. Now, we use an estimator based on a constrained $\ell_1$ minimization method proposed by \cite{p:cai-11}. Specifically, let $\widetilde{\Omega}_{1}$ be the solution to the following minimization problem:
\begin{equation*}
\min \sum_{i,j=1}^{n} |\Omega_{ij}| \quad \text{subject to} \quad \big\|\widehat{\Sigma}_{n,B}\Omega - I_n\big\|_{\infty} \leq \lambda_{n},
\end{equation*}
where, for a matrix $A = (a_{ij})_{1 \leq i,j \leq n}$, $\|A\|_{\infty} = \max_{1 \leq i,j \leq n} |a_{ij}|$ denotes the maximum norm. Here, $\lambda_{n} \in (0, \infty)$ above is the tuning parameter. 
Let $\widetilde{\Omega}_{n,B}$ be a symmetrized version of $\widetilde{\Omega}_{1}$, defined as
\begin{equation} 
\label{eq:omega}
(\widetilde{\Omega}_{n,B})_{i,j} = (\widetilde{\Omega}_{1})_{i,j} \wedge (\widetilde{\Omega}_{1})_{j,i}, \quad i,j \in \{1, \dots, n\}.
\end{equation}

Next, we require the following mild assumptions on the tail behavior of $\yunder_{1;n}$.
\begin{assumption} \label{assumption:est}
There exist $\eta \in (0,1/4)$ and $K\in (0,\infty)$ such that $\log n /B \leq \eta$ and
\begin{equation*}
\sup_{n \in \N} \max_{1\leq i \leq n} \Ex \exp\{t |y_{1,i,n}|^2\} \leq K,  \quad  t \in [-\eta,\eta].
\end{equation*}
\end{assumption}

 Note that the Gaussian innovations satisfy the above assumption. The following theorem gives a concentration result between $\widetilde{\Omega}_{n,B}$  and $\Omega_{n}$.

\begin{theorem} \label{lemma:consist}
Let $\{y_{j,t,n}\}$ be a sequence of panel time series with recursion (\ref{eq:panel}), where the common AR coefficients follow the form in (\ref{eq:rho.model}). Additionally, assume Assumption \ref{assumption:est} holds. Then, for all $\tau \in (0,\infty)$, there exists a constant $C_\tau \in (0,\infty)$ such that
\begin{equation*}
P\left( \max_{1 \leq i \leq n} |\lambda_i(\widetilde{\Omega}_{n,B}) - \lambda_i(\Omega_{n})| \leq C_{\tau} \sqrt{\frac{\log n}{B}} \right) \geq 1 - 4n^{-\tau}.
\end{equation*}
\end{theorem}
\noindent \textit{Proof}. 
From Lemma \ref{prop:explicit}, $\Omega_{n}$ is a tri-diagonal matrix, and there exists a constant $T \in (0,\infty)$ such that $\|\Omega_{n}\|_{1} = \max_{1 \leq j \leq n} \sum_{i=1}^{n} |(\Omega_{n})_{i,j}| \leq T$. Therefore, $\Omega_{n}$ belongs to the uniformity class $\mathcal{U}(q=0, s_{0}(n)=3)$, as defined in \cite{p:cai-11}, Section 3.1.

For $\tau \in (0,\infty)$, let $C_{0} = 2\eta^{-2} (2 + \tau + \eta^{-1} e^2 K^2)^2$, where $\eta$ and $K$ are parameters from Assumption \ref{assumption:est}. By applying \cite{p:cai-11}, Theorem 1(a), we obtain
\begin{equation*}
P \left( \|\widetilde{\Omega}_{n,B} - \Omega_{n} \|_{2} \leq 144 C_{0} T^{2} \sqrt{\frac{\log n}{B}} \right) \geq 1 - 4n^{-\tau},
\end{equation*}
where $\| \cdot \|_{2}$ denotes the spectral norm.

Finally, since $\max_{1 \leq i \leq n} |\lambda_i (\widetilde{\Omega}_{n,B}) - \lambda_i(\Omega_{n})| \leq \|\widetilde{\Omega}_{n,B} - \Omega_{n}\|_{2}$ due to Lemma \ref{lemma:maxineq} below, we obtain the desired result with $C_{\tau} = 144 C_{0} T^{2}$.
\hfill $\Box$

\vspace{0.5em}

From the above theorem, we conclude that $\widetilde{\Omega}_{n,B}$ is positive definite and consistently estimates $\Omega_{n}$ with high probability, provided the number of panels $B = B(n)$ is chosen such that $\lim_{n \to \infty} \log n / B(n) = 0$. Now, we proceed to estimate the outliers of $\{\Omega_n\}$ using the consistent estimator $\widetilde{\Omega}_{n,B}$ (with high probability). To accomplish this, let $\widehat{\rho}_{n}$ ($n \in \N$) be the consistent estimator of the baseline AR coefficient $\rho$ (e.g., Yule-Walker estimator). Define
\begin{equation} \label{eq:outOmega}
\widehat{out}(\widetilde{\Omega}_{n,B}) = \text{spec}(\widetilde{\Omega}_{n,B}) \cap
[a_{\widehat{\rho}_n}, b_{\widehat{\rho}_n}]^c,
\end{equation}
where $a_{\widehat{\rho}_n}$ and $b_{\widehat{\rho}_n}$ are defined as in (\ref{eq:bounds}), but replacing $\rho$ with $\widehat{\rho}_{n}$. For sets $X$ and $Y$, let
\begin{equation*}
d_{H}(X,Y) = \max\left\{ \sup_{x \in X} \inf_{y \in Y} |x-y|, \sup_{y \in Y} \inf_{x \in X} |x-y|\right\}
\end{equation*} 
be the Hausdorff distance between $X$ and $Y$.

The following theorem provides a consistent result for the outlier estimator.
\begin{theorem} \label{thm:consist}
Suppose the same set of assumptions as in Theorem \ref{lemma:consist} hold. Furthermore, assume $B = B(n)$ is such that $\lim_{n \to \infty}\log n/ B(n) = 0$. Then,
\begin{equation}
d_{H}\big( \widehat{out}(\{\widetilde{\Omega}_{n,B}\}),  out(\{\Omega_{n}\}) \big) \Pcon 0, \quad n \rightarrow \infty,
\end{equation}
where $\Pcon$ denotes convergence in probability.
\end{theorem}
\noindent \textit{Proof}. See Appendix \ref{thm:consist:proof}. \hfill $\Box$

\subsection{Numerical experiment} \label{sec:simul}

To validate our proposed outlier estimator, we conduct a simple numerical experiment. For the true model, we consider a single SCM as given in (\ref{eq:sscm}), with the length of the time series is set to $n = 100$ and the breakpoint is $k_{n} = 50$. We set the baseline AR coefficient $\rho$ to take values in $\{ 0.1, 0.3, 0.5, 0.7, 0.9\}$, and the magnitude of change $\varepsilon/\rho$ to take values in $\{ 0.5, 1, 2\}$ for each given $\rho$. For each model, we vary the panel size $B$ in $\{100, 500, 1000, 5000\}$.

For the given parameter values in the single SCM, we generate the panel time series $\{y_{j,t,n}\}$ as in (\ref{eq:panel}), where $\{z_{j,t,n}\}$ are i.i.d. standard normal random variables. Let $\Omega_{n}$ be the true precision matrix of $\underline{y}_{1;n}$, and let $\widetilde{\Omega}_{n,B}$ be its estimator as in (\ref{eq:omega}). According to Theorem \ref{thm:spacing}, the single SCM has two outliers $out(\{\Omega_{n}\}) = \{\lambda_L, \lambda_R\}$, where the explicit expressions of $\lambda_L < a_\rho < b_\rho < \lambda_R$ are provided in the same theorem. As an estimator, we use
\begin{equation*}
\widehat{out}(\widetilde{\Omega}_{n,B}) = \{ \widehat{\lambda}_L, \widehat{\lambda}_R \}, \quad
\text{where} \quad
\widehat{\lambda}_L = \lambda_{1} (\widetilde{\Omega}_{n,B}) \quad \text{and} \quad \widehat{\lambda}_R = \lambda_{n} (\widetilde{\Omega}_{n,B}).
\end{equation*}
All simulations are conducted with 1000 replications, and for each simulation, we calculate the outliers $( \widehat{\lambda}_{L,i}, \widehat{\lambda}_{R,i} )$ for $i \in \{1, \dots, 1000\}$. To assess the performance of our estimator, we compute the mean absolute error (MAE)
\begin{equation} \label{eq:mae}
\text{MAE}_{i} = \frac{1}{2} \left( |\widehat{\lambda}_{L,i} - \lambda_{L}| + |\widehat{\lambda}_{R,i} - \lambda_{R}| \right), \quad i \in \{1, \dots, 1000\},
\end{equation}
which is an equivalent norm to the Hausdorff norm.

Table \ref{tab:est} displays the average and standard deviation (in parentheses) of the MAE for each parameter setting under consideration.
\begin{table}[htbp]
    \centering
  \begin{tabular}{cc|rrrrr}
\multirow{2}{*}{$\rho$} &  \multirow{2}{*}{$\varepsilon/\rho$} & \multicolumn{4}{c}{$B$}  \\
\cline{3-7}	 
 &  & 100 & 500 & 1000 & 5000 &  \\
\hline
\multirow{4}{*}{0.1} & 0.5  & 0.15{\scriptsize (0.03)} & 0.05{\scriptsize (0.02)} & 0.03{\scriptsize (0.01)} & 0.01{\scriptsize (0.00)} \\
& 1 & 0.13{\scriptsize (0.03)} & 0.05{\scriptsize (0.02)} & 0.03{\scriptsize (0.01)} & 0.01{\scriptsize (0.01)} \\  
 & 2 & 0.10{\scriptsize (0.03)} & 0.03{\scriptsize (0.02)} & 0.03{\scriptsize (0.01)} & 0.01{\scriptsize (0.01)} \\ \hline
\multirow{3}{*}{0.3} & 0.5  & 0.09{\scriptsize(0.05)} & 0.04{\scriptsize(0.02)}& 0.02{\scriptsize(0.01)} & 0.02{\scriptsize(0.01)} \\
& 1 & 0.13{\scriptsize(0.04)} & 0.04{\scriptsize(0.02)} & 0.04{\scriptsize(0.02)} & 0.02{\scriptsize(0.01)} \\  
 & 2 & 0.22{\scriptsize(0.08)} & 0.08{\scriptsize(0.04)} & 0.06{\scriptsize(0.04)} & 0.04{\scriptsize(0.02)} \\ \hline
\multirow{3}{*}{0.5} & 0.5  & 0.20{\scriptsize(0.08)} & 0.07{\scriptsize(0.04)} & 0.07{\scriptsize(0.03)} & 0.04{\scriptsize(0.02)} \\
& 1 & 0.23{\scriptsize(0.10)} & 0.12{\scriptsize(0.07)} & 0.10{\scriptsize(0.05)} & 0.06{\scriptsize(0.03)} \\  
 & 2 &0.39{\scriptsize(0.19)} & 0.21{\scriptsize(0.11)} & 0.17{\scriptsize(0.09)} & 0.09{\scriptsize(0.04)} \\ \hline
\multirow{3}{*}{0.7} & 0.5  & 0.28{\scriptsize(0.15)} & 0.08{\scriptsize(0.04)} & 0.04{\scriptsize(0.02)} & 0.03{\scriptsize(0.02)} \\
& 1 & 0.39{\scriptsize(0.20)} & 0.09{\scriptsize(0.05)} & 0.07{\scriptsize(0.05)} & 0.05{\scriptsize(0.03)} \\  
 & 2 & 0.67{\scriptsize(0.31)} & 0.17{\scriptsize(0.10)} & 0.15{\scriptsize(0.10)} & 0.10{\scriptsize(0.06)} \\ \hline
\multirow{3}{*}{0.9} & 0.5  & 0.48{\scriptsize(0.22)} & 0.13{\scriptsize(0.07)} & 0.07{\scriptsize(0.05)} & 0.04{\scriptsize(0.03)} \\
& 1 & 0.55{\scriptsize(0.29)} & 0.14{\scriptsize(0.08)} & 0.09{\scriptsize(0.07)} & 0.05{\scriptsize(0.04)} \\  
 & 2 & 0.54{\scriptsize(0.33)} & 0.26{\scriptsize(0.18)} & 0.18{\scriptsize(0.12)} & 0.07{\scriptsize(0.06)} \\ \hline
\end{tabular} 

\caption{The average and standard deviation (in parentheses) of the mean absolute error of the single SCM for each combination of $(\rho, \varepsilon, B)$. }
\label{tab:est}
\end{table}

In all simulations, as the panel size $B$ increases, the MAE decreases and tends to zero. Moreover, the finite sample bias of $\widehat{\lambda}_{L}$ and $\widehat{\lambda}_{R}$ due to $n$ and $k$ is negligible for moderate baseline AR coefficients (i.e., $\rho \in \{0.1, 0.3\}$) and reasonably small for larger $\rho \in \{0.5, 0.7, 0.9\}$. These observations align with Remark \ref{rmk:spacing}(ii), which states that the bias due to the breakpoint $k$ is bounded by $c^k$ for some $c \in (0,1)$, where $c = c(\rho)$ approaches one as $|\rho|$ approaches one, leading to a larger bias. However, the effect of the magnitude of change $\varepsilon$ is not consistent. For $\rho \in \{0.3, 0.5, 0.7, 0.9\}$, the bias tends to increase as $\varepsilon/\rho$ increases. In contrast, an opposite trend is observed for $\rho = 0.1$.

\input{proofs-main}

\section{Concluding remarks and extensions} \label{sec:conclusion}

In this article, we study structural changes in nonstationary AR$(1)$ processes when the change period of the AR coefficients is very short. Our approach focuses on the edge eigenvalues of the precision matrix of the observed time series. We show that under the null hypothesis of no structural change, all eigenvalues of the precision matrix lie within the bulk spectrum, with the distribution explicitly described in (\ref{eq:cdf}). Conversely, under the alternative hypothesis of structural change, we show that there are outliers (edge eigenvalues) that deviate from the bulk spectrum observed in the null case.

We also propose a consistent estimator for the outliers (under both null and SCM scenarios) within the panel time series framework. Unlike traditional CUSUM and disturbed ARMA models, which make estimations and inferences based on a single realization of the time series, our proposed method requires the panel size $B = B(n)$ to satisfy $\log n \ll B(n) \ll n$ as $n \rightarrow \infty$. This requirement arises because traditional methods assume $\tau_j > 0$ in (\ref{eq:trad}), relying on sufficiently large observations with structural changes, whereas our statistical model assumes $\tau_j = 0$.

By using the dichotomy in Theorem \ref{thm:1} and (\ref{eq:outSCM}), one can examine the hypotheses $H_0: \text{out}(\Omega_{n}) = \emptyset$ versus $H_a: \text{out}(\Omega_{n}) \neq \emptyset$ and utilize $\widehat{\text{out}}(\widetilde{\Omega}_{n,B})$ as a test statistic to formulate tests for structural change within the panel time series framework. However, deriving the asymptotic behavior of $\widehat{\text{out}}(\widetilde{\Omega}_{n,B})$ under both the null and SCM scenarios is a highly nontrivial task, making direct calculation of the $\alpha$-level critical region infeasible. Therefore, alternative approaches, such as bootstrap methods to construct confidence intervals, could be useful in this context. This will be investigated in future studies.

Lastly, the problem under consideration in this article is specific to AR processes of order one. We discuss the possible extension of our results to nonstationary AR processes of general order $p \in \mathbb{N}$. To do so, similar to (\ref{eq:tri-AR}), suppose we observe $\underline{y}_n = (y_{1,n}, \dots, y_{n,n})^\top$ with recursion $y_{t,n} = \sum_{j=1}^{p} \phi_{t,n}^{(j)} y_{t-j,n} + z_{t,n}$, $n \in \mathbb{N}$, $t \in \{1, \dots, n\}$. Analogous to (\ref{eq:rho.model}), we assume $\underline{\phi}_{t,n} = (\phi_{t,n}^{(1)}, \dots, \phi_{t,n}^{(p)})^\top \in \mathbb{R}^{p}$ satisfies
\begin{equation*}
\underline{\phi}_{t,n} = \underline{\phi} + \sum_{j=1}^{m} \underline{\varepsilon}_j I_{E_{j,n}}(t), \quad t \in \{1, \dots, n\}.
\end{equation*}
Here, the baseline AR$(p)$ coefficient $\underline{\phi} = (\phi_1, \dots, \phi_p)^\top$ is such that $\phi(z) = 1 - \sum_{j=1}^{p} \phi_j z^{j}$ does not have zeros on or inside the unit circle. Then, using similar techniques to those used in the proof of Theorem \ref{thm:1}, one can show that
\begin{equation*}
\mu_{A_n} \xrightarrow{D} \mu_{\underline{\phi}} \quad \text{as} \quad n \rightarrow \infty,
\end{equation*}
provided (\ref{eq:Ejn-0}) holds. Here, the LSD $\mu_{\underline{\phi}}$ is the probability measure on $\mathbb{R}$ with distribution
\begin{equation*}
F_{\mu_{\underline{\phi}}}(t) = \mu_{\underline{\phi}}((-\infty, t]) = \frac{1}{2\pi} \int_{0}^{2\pi} I_{(-\infty, t]}(|\phi(e^{-ix})|^2) \, dx, \quad t \in \mathbb{R}.
\end{equation*}

However, generalizing the outlier results stated in Section \ref{sec:single} to higher-order AR processes, even for $p=2$, seems challenging. This difficulty arises because, when calculating the outliers for the AR$(1)$ SCM, we need to determine the explicit expression of $\lim_{n \to \infty} M_{n,r} = M_r$, where $M_{n,r}$ is defined as in (\ref{eq:M_nr}). Deriving expressions for $M_r$ when $p=1$ involves detailed calculations, and a general expression for $M_r$ for $p \in \mathbb{N}$ and $\{\underline{\varepsilon}_j\}_{j=1}^{m}$ appears intractable. However, we conjecture that for nonstationary AR$(p)$ models, Theorem \ref{lemma:outlier.null} holds under the null hypothesis of no structural change, and outliers are observed when structural change occurs. This will also be investigated in future studies.



\appendix \label{appen}

\input{asd}
\input{proofs}

\input{technical_detail}

\section*{Acknowledgement} \label{sec:acknowledgement}
The author's research was supported by Taiwan's National Science and Technology Council (grant 110-2118-M-001-014-MY3). The research was partially conducted while the author was affiliated with Texas A\&M University and was partly supported by NSF grant DMS-1812054. The author acknowledges Professor Suhasini Subba Rao for fruitful discussions. Additionally, the author wishes to thank the two anonymous referees and editors for their valuable comments and corrections, which have significantly enhanced the quality of the article in all aspects.

\bibliographystyle{plainnat}
\bibliography{bib-ar1}

\end{document}

%% file: proofs-main.tex
\section{Proof of Theorem \ref{thm:spacing}} \label{thm:spacing:proof}

This section contains the proof of Theorem \ref{thm:spacing}. For brevity, we primarily focus on the case $\rho \in (0,1)$. The proof for $\rho \in (-1,0)$ can be treated similarly. The proof strategy is motivated by Theorem 2.1 of \cite{p:ben-nad-11}. To prove the theorem, it suffices to show the following four statements:

\vspace{0.3em}

\noindent $\bullet$ In the case when $|\rho| \leq |\rho + \varepsilon|$, there exists $c \in (0,1)$ such that for any fixed $j \in \mathbb{N}$:
    \begin{itemize}
        \item[(A)] $\lambda_1(B_n)  = M + O(c^{k})$, as $n \rightarrow \infty$.
        \item[(B)] $\lambda_n(B_n)  = m + O(c^{k})$, as $n \rightarrow \infty$.
        \item[(C)] $\lim_{n \to \infty} \lambda_{j+1} (B_n) = b_\rho$ and $\lim_{n \to \infty} \lambda_{n-j}(B_n) = a_\rho$.
    \end{itemize}

\vspace{0.3em}

\noindent $\bullet$ In the case when $|\rho| > |\rho + \varepsilon|$, for any fixed $j \in \mathbb{N}$:
    \begin{itemize}
        \item[(D)] $\lim_{n \to \infty} \lambda_{j} (B_n) = b_\rho$ and $\lim_{n \to \infty} \lambda_{n+1-j}(B_n) = a_\rho$.
    \end{itemize}

The proof of (A)–(D) above can be divided into seven steps, which we will briefly summarize below:

\begin{itemize}
    \item[1.] We show that there are at most two outliers in the sequence $\{B_n\}$, one on the left and one on the right of the bulk spectrum.

    \item[2.] Using spectral decomposition, we obtain a $3 \times 3$ matrix, where the zeros of the determinant of this matrix represent the potential outliers.

    \item[3.] We show that the matrix from Step 2 is a block diagonal matrix with block sizes one and two.

    \item[4.] We show that the $1 \times 1$ block from Step 3 does not have a zero within the possible range for outliers. Therefore, the potential outliers are the zeros of the determinant of the $2 \times 2$ submatrix.

    \item[5.] We prove that if $|\rho| > |\rho+\varepsilon|$, then there are no zeros of the determinant of the $2 \times 2$ submatrix within the possible range, implying no outliers exist.

    \item[6.] We show that if $|\rho| < |\rho+\varepsilon|$, then there are exactly two zeros, and we derive their explicit forms.

    \item[7.] We calculate the approximation errors due to the breakpoint $k$ to complete the proof.
\end{itemize}

Now, we provide detailed explanations for each step.

\subsubsection*{Step 1}
Let $A_{0,n}$ and $B_n$ be the precision matrices under the null and the single SCM, respectively, and let $\widetilde{A}_{0,n}$ be defined as in (\ref{eq:perturb}). Define $P_n = B_n - A_{0,n}$ and $\widetilde{P}_n = B_n - \widetilde{A}_{0,n}$. By using Lemma \ref{prop:explicit}, we have
\begin{equation} \label{eq:pform}
[P_n]_{i,j} =  
\left\{
\begin{array}{ll}
\varepsilon(\varepsilon + 2\rho), & (i,j) = (k-1,k-1), \\
-\varepsilon, & (i,j) \in \{(k-1,k), (k,k-1)\}, \\
0, & \text{otherwise}.
\end{array} 
\right.
~~\text{and}~~
\widetilde{P}_n = P_n - \rho^2 E_n,
\end{equation}
where $E_n = \diag(0, \dots, 0, 1) \in \R^{n \times n}$. Therefore, provided that $\varepsilon \neq 0$, $P_n$ has exactly two nonzero eigenvalues, which are the solutions of the quadratic equation
\begin{equation*}
z^2 - (\varepsilon^2 + 2\rho \varepsilon) z - \varepsilon^2 = 0.
\end{equation*}
We denote these two nonzero eigenvalues as $\alpha < \beta$. Since $\alpha \beta = -\varepsilon^2 < 0$, we have $\alpha < 0 < \beta$. Therefore, the eigenvalues of $P_n$ are given by $\lambda_1(P_n) = \beta$, $\lambda_n(P_n) = \alpha$, and $\lambda_2(P_n) = \dots = \lambda_{n-1}(P_n) = 0$. Next, by applying Lemma \ref{lemma:weyl} below, we have
\begin{equation*}
\lambda_{j+1}(A_{0,n}) \leq \lambda_j(B_n) \leq \lambda_{j-1}(A_{0,n}), \quad j \in \{2, \dots, n-1\}.
\end{equation*}
Therefore, by Lemma \ref{lemma:tight} and the sandwich property, we have
\begin{equation*}
\lim_{n \to \infty} \lambda_{j+1}(B_n) = b_\rho \quad \text{and} \quad
\lim_{n \to \infty} \lambda_{n-j}(B_n) = a_\rho, \quad j \in \N.
\end{equation*}
This proves (C) and part of (D). Next, by Theorem \ref{prop:asymdist}, since $\mu_{B_n} \xrightarrow{d} \mu_\rho$ as $n \rightarrow \infty$, we conclude that the possible outliers of $\{B_n\}$ are the limiting points of $\lambda_1(B_n)$ or $\lambda_n(B_n)$.

\subsubsection*{Step 2}
Let $\widetilde{A}_{0,n} = U_n \Lambda_n U_n^{\top}$ be an eigen-decomposition, where $U_n = (u_1^{(n)}, \dots, u_n^{(n)} )$ is an orthonormal matrix and $\Lambda_n = \diag (\lambda_1(\widetilde{A}_{0,n}), \dots, \lambda_n(\widetilde{A}_{0,n}))$ is a diagonal matrix (explicit expressions of $U_n$ and $\Lambda_n$ are given in (\ref{eq:evalperturb}) and (\ref{eq:evecperturb}), respectively). For the sake of convenience, we omit the index $n$ and write $u_i = u_i^{(n)}$ and $\lambda_i = \lambda_i(\widetilde{A}_{0,n})$.

Next, let $\widetilde{P}_n = V_n \Theta_r V_n^{\top}$ be a spectral decomposition of $\widetilde{P}_n$, where $r$ is the rank of $\widetilde{P}_n$, $\Theta_r$ is a diagonal matrix of the nonzero eigenvalues of $\widetilde{P}_n$, and $V_n$ is an $n \times r$ matrix with columns consisting of $r$ orthogonal eigenvectors. Since the explicit form of $\widetilde{P}_n$ is given in (\ref{eq:pform}), we can fully determine the elements in the spectral decomposition of $\widetilde{P}_n$ by
\begin{equation}
\label{eq:ptildeform}
r=3, \quad V_n = ( a_1 e_{k-1} + b_1 e_{k}, a_2 e_{k-1} + b_2 e_{k}, e_n ), \quad \text{and} \quad \Theta_r = \diag(\theta_1, \theta_2, \theta_3).
\end{equation}
Here, $(a_{1}, b_{1})^{\top}$ and $(a_{2}, b_{2})^{\top}$ are the two orthonormal eigenvectors of the matrix 
\begin{equation} \label{eq:Erho}
\varepsilon \begin{pmatrix}
\varepsilon + 2\rho & -1 \\
-1 & 0 \\
\end{pmatrix},
\end{equation}
$e_{k}$ is the $k$th canonical basis of $\mathbb{R}^{n}$, and $\theta_i$, $i\in \{1,2,3\}$, are given by 
\begin{equation*}
\theta_1 = \varepsilon \frac{(\varepsilon + 2\rho) - \sqrt{ (\varepsilon+2\rho)^2 + 4 } }{2},\quad 
\theta_2 = \varepsilon \frac{(\varepsilon + 2\rho) + \sqrt{ (\varepsilon+2\rho)^2 + 4 } }{2}, \quad 
\theta_3 = - \rho^2.
\end{equation*}
We note that the eigenvalues corresponding to the eigenvectors $(a_{1}, b_{1})^{\top}$ and $(a_{2}, b_{2})^{\top}$ of (\ref{eq:Erho})
are $\theta_1$ and $\theta_2$, respectively.

Since $U_n$ is orthogonal, we have
\begin{equation*} 
B_n = \widetilde{A}_{0,n} + \widetilde{P}_n = U_n \big( \Lambda_n + U_n V_n \Theta_r V_n^{\top} U_n^{\top} \big) U_n^{\top}.
\end{equation*} 
Therefore, $\text{spec}(B_n) = \text{spec}(\Lambda_n + S_n \Theta_r S_n^{\top})$, where 
\begin{equation}
\label{eq:S_n}
S_n = U_n V_n = ( a_1 u_{k-1} + b_1 u_{k}, a_2 u_{k-1} + b_2 u_{k}, u_n ).
\end{equation}
Next, by using \cite{p:arb-88}, Theorem 2.3, $z \in [a_\rho, b_\rho]^c \setminus \{ \lambda_1, \dots, \lambda_n \}$ is an eigenvalue of $B_n$ (thus, an eigenvalue of $\Lambda_n + S_n \Theta_r S_n^{\top}$) if and only if 
\begin{equation}
\det \left( I_r - S_n^\top (zI_n - \Lambda_n)^{-1} S_n \Theta_r \right) = 0.
\end{equation}
Therefore, we conclude that $z$ is an eigenvalue of $B_n$ but not of $\widetilde{A}_{0,n}$ if and only if the matrix 
\begin{equation}
\label{eq:M_nr}
M_{n,r} = I_r - S_n^{\top} (zI_n - \Lambda_n)^{-1} S_n \Theta_r \in \mathbb{R}^{3 \times 3}
\end{equation}
is singular.

\subsubsection*{Step 3}
The $(i,j)$th ($i,j \in \{1,2,3\}$) component of $M_{n,r}$ is given by
\begin{equation}
\begin{aligned}
[M_{n,r}]_{i, j} &= \delta_{i=j} - \sum_{\ell = 1}^{n} [S_n^{\top}]_{i, \ell} [(zI_n - \Lambda_n)^{-1}]_{\ell, \ell} [S_n]_{\ell, j} [\Theta_r]_{j,j} \\
&= \delta_{i=j} - \theta_j \sum_{\ell = 1}^{n} \frac{[S_n]_{\ell, i} [S_n]_{\ell, j}}{z - \lambda_{\ell} (\widetilde{A}_{0,n})} \\
&= \delta_{i=j} - \theta_j \sum_{\ell = 1}^{n} \frac{[S_n]_{\ell, i} [S_n]_{\ell, j}}{2\rho \cos \left( \frac{\ell \pi}{n+1} \right) + \left( z - (1+\rho^2) \right)}.
\end{aligned}
\label{eq:Mmatrix}
\end{equation}
We make an approximation of the above term. Let $a = \frac{z - (1+\rho^2)}{2\rho}$. Therefore, if $z > b_\rho = (1+\rho)^2$, then $a > 1$; if $z < (1-\rho)^2$, then $a < -1$. From (\ref{eq:evecperturb}), we have
\begin{eqnarray*}
&& \sum_{\ell = 1}^{n} \frac{[S_n]_{\ell, 1} [S_n]_{\ell, 3}}{\cos \left( \frac{\ell \pi}{n+1} \right)+ a } \nonumber \\ 
&&~= 
\frac{2a_1}{n+1} \sum_{\ell = 1}^{n} \frac{ \sin \left(\frac{n \ell \pi}{n+1}\right) \sin \left(\frac{k \ell \pi}{n+1}\right) }{\cos \left(\frac{\ell \pi}{n+1}\right) + a } + \frac{2b_1}{n+1} \sum_{\ell = 1}^{n} \frac{ \sin \left(\frac{n \ell \pi}{n+1}\right) \sin \left(\frac{(k+1) \ell \pi}{n+1}\right) }{\cos \left( \frac{\ell \pi}{n+1}\right) + a } \nonumber \\
&&~= \frac{2a_1}{n+1} \sum_{\ell = 1}^{n} \frac{ (-1)^{\ell+1} \sin \left(\frac{\ell \pi}{n+1}\right) \sin \left(\frac{k \ell \pi}{n+1}\right) }{\cos \left(\frac{\ell \pi}{n+1}\right) + a } + \frac{2b_1}{n+1} \sum_{\ell = 1}^{n} \frac{(-1)^{\ell+1} \sin \left(\frac{\ell \pi}{n+1}\right) \sin \left(\frac{(k+1) \ell \pi}{n+1}\right) }{\cos \left(\frac{\ell \pi}{n+1}\right) + a },
\end{eqnarray*}
where $a_1$ and $b_1$ are from (\ref{eq:ptildeform}). Therefore, by using Lemma \ref{lemma:int} below, we have
\begin{eqnarray*}
\lim_{n \rightarrow \infty} \frac{2}{n+1} \sum_{\ell = 1}^{n} \frac{ (-1)^{\ell+1} \sin \left(\frac{\ell \pi}{n+1}\right) \sin \left(\frac{k \ell \pi}{n+1}\right) }{\cos \left(\frac{\ell \pi}{n+1}\right) + a } &=& \lim_{n \rightarrow \infty} \frac{2\pi}{n+1} \sum_{\ell \text{: odd}}^{n} \frac{\sin \left(\frac{\ell \pi}{n+1}\right) \sin \left(\frac{k \ell \pi}{n+1}\right) }{\cos \left(\frac{\ell \pi}{n+1}\right) + a } \\
&&- \lim_{n \rightarrow \infty} \frac{2\pi}{n+1} \sum_{\ell \text{: even}}^{n} \frac{ \sin \left(\frac{\ell \pi}{n+1}\right) \sin \left(\frac{k \ell \pi}{n+1}\right) }{\cos \left(\frac{\ell \pi}{n+1}\right) + a } \\
&=& \frac{1}{2}\left( G(1,k) - G(1,k)\right) = 0,
\end{eqnarray*} where $G(1,k)$ is from Lemma \ref{lemma:int} below.
This indicates that $\lim_{n \rightarrow \infty} [M_{n,r}]_{1,3} = 0$. Let $M_r = \lim_{n \rightarrow \infty} M_{n,r}$. Then, by using similar calculations with the help of Lemma \ref{lemma:int}, $M_r$ can be written as
\begin{equation*}
M_r = \begin{pmatrix}
p & q & 0 \\
q & r & 0 \\
0 & 0 & A \\
\end{pmatrix},
\end{equation*}
for some $p, q, r, A$ which we will elaborate on in the next step. Therefore, the singularities of $M_r$ come from either solving $A = 0$ or solving $pr - q^2 = 0$.

\subsubsection*{Step 4}
First, we will calculate $A$. Using Lemma \ref{lemma:int}, we have
\begin{eqnarray*}
\lim_{n \rightarrow \infty} \sum_{\ell = 1}^{n} \frac{[S_n]_{\ell, 3} [S_n]_{\ell, 3}}{\cos \frac{\ell \pi}{n+1} + a } &=& \lim_{n \rightarrow \infty} \frac{2}{n+1} \sum_{\ell = 1}^{n} \frac{ \sin^2 \frac{\ell \pi}{n+1} }{\cos \frac{\ell \pi}{n+1} + a } \\
&=& \frac{2}{\pi} \int_{0}^{\pi} \frac{\sin^2 x}{\cos x + a} \, dx = \frac{1}{\pi} \int_{0}^{2\pi} \frac{\sin^2 x}{\cos x + a} \, dx = G(1,1).
\end{eqnarray*}
Therefore, from (\ref{eq:Mmatrix}), we have $A = 1 - \theta_3 G(1,1)/(2\rho) = 1 + \rho G(1,1)/2$.
Let $\rho \in (0,1)$ and let $z < a_\rho$. Since $z < a_\rho$, we have $a = \{z - (1+\rho^2)\}/(2\rho) \in (-\infty, -1)$. Let
\begin{equation*}
z_1 = -a - \sqrt{a^2 - 1} \quad \text{and} \quad z_2 = -a + \sqrt{a^2 - 1}.
\end{equation*}
Then, by Lemma \ref{lemma:int} again, we have
\begin{equation*}
G(1,1) = \frac{2}{z_1 - z_2} (1 - z_1^2) = -\frac{1}{\sqrt{a^2 - 1}} \left( 1 - \left(-a - \sqrt{a^2 - 1}\right)^2 \right) = 2 \left(a + \sqrt{a^2 - 1}\right).
\end{equation*}
This indicates that $G(1,1)$ is a decreasing function of $a$ on the domain $(-\infty, -1)$, thus $G(1,1) > -2$ and $A = 1 +\rho G(1,1)/2 > 1 - \rho > 0$.
Therefore, we conclude that there is no $z \in (-\infty, a_\rho)$ such that $A(z) = 0$. Similarly, we can show that there is no $z \in (b_\rho, \infty)$ such that $A(z) = 0$.

By using similar techniques with help of Lemma \ref{lemma:int}, elements in $[M_r]_{1,1}$, $[M_r]_{2,2}$, and $[M_r]_{1,2}$ are given by
\begin{eqnarray*}
&& \lim_{n \rightarrow \infty} \sum_{\ell = 1}^{n} \frac{[S_n]_{\ell, 1} [S_n]_{\ell, 1}}{\cos \frac{\ell \pi}{n+1} + a } \\
 && \quad = \frac{2a_1^2}{\pi} \int_{0}^{\pi} \frac{ \sin (kx) \sin (kx) }{\cos x + a } dx + \frac{2b_1^2}{\pi} \int_{0}^{\pi} \frac{ \sin ((k+1)x) \sin ((k+1)x) }{\cos x + a } dx \\
&&\quad ~~+ \frac{2a_1b_1}{\pi} \int_{0}^{\pi} \frac{ \sin (kx) \sin ((k+1)x) }{\cos x + a } dx \\
&& \quad = a_1^2 G(k,k) + b_1^2 G(k+1,k+1) + 2a_1b_1 G(k,k+1), \\
&& \lim_{n \rightarrow \infty} \sum_{\ell = 1}^{n} \frac{[S_n]_{\ell, 2} [S_n]_{\ell, 2}}{\cos \frac{\ell \pi}{n+1} + a } = a_2^2 G(k,k) + b_2^2 G(k+1,k+1) + 2a_2b_2 G(k,k+1), \\
&& \lim_{n \rightarrow \infty} \sum_{\ell = 1}^{n} \frac{[S_n]_{\ell, 1} [S_n]_{\ell, 2}}{\cos \frac{\ell \pi}{n+1} + a } = a_1a_2 G(k,k) + b_1b_2 G(k+1,k+1) + (a_1b_2+a_2b_1) G(k,k+1).
\end{eqnarray*}
Furthermore, for $k \rightarrow \infty$, we have
\begin{equation}
\begin{aligned}
\frac{1}{2\rho} G(k,k) &\approx \frac{|a|}{a} \frac{1}{\sqrt{a^2-1}} = \frac{|a|}{a} \frac{1}{\sqrt{ (z-a_\rho) (z-b_\rho) }} =: G(z),\\
 \frac{1}{2\rho} G(k,k+1) &\approx \frac{z_2}{2 \rho (z_2 - z_1)} = \widetilde{G}(z).
\end{aligned}
\label{eq:Gapprox}
\end{equation}
The approximation errors above are of order $O\left( |z_{1}|^k \wedge |z_{2}|^{k}\right)$ as $k \rightarrow \infty$.
By (\ref{eq:Gapprox}) and Lemma \ref{lemma:eigenvalues}(a,b), the $2 \times 2$ submatrix of $M_{r}$ limits to
\begin{equation}
\label{eq:2by2approx}
 \widetilde{M}_{r-1} = 
\begin{pmatrix}
1 - \theta_1 \left( G(z) + 2a_1b_1 \widetilde{G}(z) \right) & \theta_2 (a_1b_2 + a_2b_1) \widetilde{G}(z) \\
\theta_1 (a_1b_2 + a_2b_1) \widetilde{G}(z) & 1 - \theta_2 \left( G(z) + 2a_2b_2 \widetilde{G}(z) \right).
\end{pmatrix}
\end{equation}
Therefore, we have
\begin{eqnarray*}
\det \widetilde{M}_{r-1} &=& (1-\theta_1 G) (1-\theta_2 G) -2 \big( (1-\theta_2G) \theta_1 a_1 b_1 + (1-\theta_1 G) \theta_2a_2b_2 \big) \widetilde{G}  \\
&&+ 4 \theta_1 \theta_2 a_1a_2b_1b_2 \widetilde{G}^2 -\theta_1\theta_2 (a_1b_2+a_2b_1)^2 \widetilde{G}^2.
\end{eqnarray*}
Furthermore, by using Lemma \ref{lemma:eigenvalues}(b)--(g), $\det \widetilde{M}_{r-1}$ can be further simplified as
\begin{equation}
\begin{aligned}
\det \widetilde{M}_{r-1} &=
 (1-\theta_1 G) (1-\theta_2 G) -2 \big( \theta_1 a_1 b_1 + \theta_2a_2b_2 \big) \widetilde{G} - \theta_1\theta_2 \widetilde{G}^2 \\
&= (1-\theta_1 G) (1-\theta_2 G) + 2 \varepsilon \widetilde{G} + \varepsilon^2 \widetilde{G}^2. 
\end{aligned}
\label{eq:Msol}
\end{equation}
We note that since we use an approximation in (\ref{eq:Gapprox}), the exact determinant of the submatrix of $M_r$ (which is $pr - q^2$) is not identical to $\det \widetilde{M}_{r-1}$. However, by using the approximation error as in (\ref{eq:sineint1}), we have
\begin{equation}
\label{eq:Msolapprox}
pr - q^2 = \det \widetilde{M}_{r-1} + O\left( \frac{z_2^{4k+2}}{(1 - z_2^2)^2} \right), \quad k \rightarrow \infty.
\end{equation}

\subsubsection*{Step 5}
We first show that if $|\rho + \varepsilon| < |\rho|$, then $\det \widetilde{M}_{r-1}$ does not have zeros in $[a_\rho, b_\rho]^{c}$. Let $\rho \in (0,1)$, so $\varepsilon \in (-\rho, 0)$. Moreover, we only consider the case when $z > b_\rho$, or equivalently $a > 1$. The case when $z < a_\rho$ can be treated similarly. By using (1) $\theta_1 + \theta_2 = \varepsilon^2 + 2\rho \varepsilon$ and (2) $\theta_1 \theta_2 = -\varepsilon^2$, we have
\begin{equation}
\label{eq:Mepsol}
f_z(\varepsilon) = \det \widetilde{M}_{r-1} = (\widetilde{G}^2 - G^2 - G) \varepsilon^2 + 2 (\widetilde{G} - \rho G) \varepsilon + 1.
\end{equation}
Recall
\begin{equation*}
G = \frac{2}{2\rho(z_2 - z_1)} = \frac{1}{\sqrt{(z - a_\rho) (z - b_\rho)}} \text{~~and~~}
\widetilde{G} = \frac{2 z_2}{2\rho(z_2 - z_1)} = \frac{z_2}{\sqrt{(z - a_\rho) (z - b_\rho)}}.
\end{equation*}
Since $z_2 \in (-1,0)$, we have $\widetilde{G}(z) < 0 < -\widetilde{G}(z) < G(z)$. Therefore, the leading coefficient of $f_z(\varepsilon)$ is negative. Since $f_z(0) > 0$, we conclude
\begin{eqnarray*}
&& f_z(\cdot) \text{ does not have a solution in } \varepsilon \in (-\rho, 0) \\
&&~~~  \Longleftrightarrow f_z(-\rho) = (\widetilde{G}^2 - G^2 - G) \rho^2 - 2(\widetilde{G} - \rho G) \rho + 1 > 0, \quad z > b_\rho.
\end{eqnarray*}


Next, we parametrize $z_2 = \cos x$ for $x \in (\pi/2, \pi)$ (since $z_2 \in (-1,0)$). For simplicity, let $C = \cos x$ and $S = \sin x$. Then, we have
\begin{equation} \label{eq:trig1}
G = \frac{2}{2\rho(z_2 - 1/z_2)} = -\frac{C}{\rho S^2} \text{~~and~~} 
\widetilde{G} = \frac{2 z_2}{2\rho(z_2 - 1/z_2)} = -\frac{C^2}{\rho S^2}.
\end{equation}
Substitute (\ref{eq:trig1}) into $f_z(-\rho)$ and multiply by $S^2$, we get
\begin{equation*}
S^2 f_z(-\rho) = -C^2 + \rho C + 2C^2 + S^2 = 1 + \rho C > 0.
\end{equation*}
Therefore, when $|\rho + \varepsilon| < |\rho|$, there is no solution for $\det \widetilde{M}_{r-1} = 0$, thus we conclude that the eigenvalues of $B_n$ do not have outliers. This completes the proof of (D).

\subsubsection*{Step 6}
Consider the case $|\rho| < |\rho + \varepsilon|$ and assume $0 < \rho < \rho + \varepsilon$. The case when $\rho + \varepsilon < \rho < 0$ can be treated similarly. Now, we obtain the solution of (\ref{eq:Mepsol}) in two different regions. 

First, we assume $z \in (b_\rho, \infty)$. Then, using the same parametrization as in (\ref{eq:trig1}), we have
\begin{equation}  \label{eq:trig2}
\begin{aligned}
&f_{z}(\varepsilon) = \left( \frac{C^4}{\rho^2 S^4} - \frac{C^2}{\rho^2 S^4} + \frac{C}{S^2} \right) \varepsilon^2 + 2 \left( -\frac{C^2}{\rho S^2} + \frac{C}{S^2} \right) \varepsilon + 1 = 0 \\
& \quad \Longleftrightarrow -(\varepsilon + \rho)^2 C^2 + \varepsilon \rho (\varepsilon + 2\rho) C + \rho^2 = 0.
\end{aligned}
\end{equation}
Since $\varepsilon(\varepsilon+2 \rho) >0$ (here we use the condition $0<\rho<\rho+\varepsilon$), solution $C \in(-1,0)$ of (\ref{eq:trig2}) is given by 
\begin{equation*}
C = z_2 =  \frac{ \rho \varepsilon (\varepsilon+2\rho) - \sqrt{ \rho^2 \varepsilon^2 (\varepsilon+2\rho)^2 + 4\rho^2 (\varepsilon +\rho)^2 }}{2 (\varepsilon + \rho)^2}.
\end{equation*}
Recall $z_2 = -a + \sqrt{a^2 -1}$ and $a = \{z- (1+\rho^2)\}/(2\rho)$. Thus, the solution $z \in (b_\rho, \infty)$ is 
\begin{equation*}
\mathfrak{M} = 1+ \rho^2 - \rho \left( C+C^{-1} \right).
\end{equation*}
Since $\varepsilon(\varepsilon+2 \rho) > 0$ (using the condition $0 < \rho < \rho + \varepsilon$), the solution$C \in (-1, 0)$ of (\ref{eq:trig2}) is given by 
\begin{equation*} 
C = z_2 =  \frac{ \rho \varepsilon (\varepsilon + 2\rho) - \sqrt{ \rho^2 \varepsilon^2 (\varepsilon + 2\rho)^2 + 4\rho^2 (\varepsilon + \rho)^2 }}{2 (\varepsilon + \rho)^2}.
\end{equation*}
Recall $z_2 = -a + \sqrt{a^2 -1}$ and $a = \{z - (1 + \rho^2)\}/(2\rho)$. Thus, the solution $z \in (b_\rho, \infty)$ is 
\begin{equation*}
\mathfrak{M} = 1 + \rho^2 - \rho \left( C + C^{-1} \right).
\end{equation*}

Second, we assume $z \in (-\infty, a_\rho)$. In this case, we have $a = \{z - (1+\rho^2)\}/(2\rho) \in (-\infty, -1)$, thus
\begin{equation*}
G = \frac{2}{2\rho(z_1 - z_2)} = \frac{-1}{\sqrt{(z-a_\rho) (z-b_\rho)}}\text{~~and~~}
\widetilde{G} = \frac{2 z_1}{2\rho(z_1 - z_2)} = \frac{-z_1}{\sqrt{(z-a_\rho) (z-b_\rho)}}.
\end{equation*}
Since $0 < z_1 < 1 < z_2$, we parametrize $z_1 = \cos x = C^\prime$ for some $x \in (0, \pi/2)$. Then, equation (\ref{eq:trig2}) remains the same, but our solution is in $(0,1)$. Thus,
\begin{equation*} 
C^\prime = z_1 = \frac{\rho \varepsilon (\varepsilon+2\rho) + \sqrt{\rho^2 \varepsilon^2 (\varepsilon+2\rho)^2 + 4\rho^2 (\varepsilon +\rho)^2 }}{2 (\varepsilon + \rho)^2}.
\end{equation*}
Using a similar argument as above, the solution for $z \in (-\infty, a_\rho)$ is
\begin{equation*}
\mathfrak{m} = 1 + \rho^2 - \rho \left( C^{\prime} + (C^{\prime})^{-1} \right).
\end{equation*} 
To conclude, when $|\rho| < |\rho+\varepsilon|$, we show that there are exactly two outliers: one on the right ($=\mathfrak{M}$) and the other on the left ($=\mathfrak{m}$). This proves (A).

\subsubsection*{Step 7}
Lastly, we consider the effect of the break point $k$ and prove Remark \ref{rmk:spacing}(ii).
Let $X(z) = -(\varepsilon + \rho)^2 z^2 + \varepsilon \rho (\varepsilon + 2\rho) z + \rho^2$. Then, it is easy to check that $X(1) < 0 < X(\rho)$.
Therefore, there exists $\widetilde{z} \in (\rho, 1)$ such that $X(\widetilde{z}) = 0$. Moreover, it is easy to check that $\widetilde{z}$ is not a multiple root. Therefore, around $z=\widetilde{z}$, $X(z)$ changes its sign. By (\ref{eq:Msolapprox}),
\begin{equation*}
(pr-q^{2})(\widetilde{z}) = \underbrace{X(\widetilde{z})}_{=0} + O\left( \frac{ (\widetilde{z})^{4k+2} }{(1-(\widetilde{z})^2)^2} \right).
\end{equation*}
Therefore, for a sufficiently large $k$, there exists $c \in (0,1)$ and an interval $I(\widetilde{z}) = [\widetilde{z}-c^{k}, \widetilde{z}+c^{k}] \subset (-1,0)$ such that $(pr-q^{2})(z)$ has a zero in $I(\widetilde{z})$. Let the solution be $\widehat{z}$. Then, $\widehat{\mathfrak{m}} = 1 + \rho^2 - \rho(\widehat{z} + \widehat{z}^{-1})$ is the ``true'' outlier, and $\mathfrak{m} = 1 + \rho^2 - \rho(\widetilde{z} + \widetilde{z}^{-1})$ is an approximate solution as described in [Step 6]. Since $|\widehat{z} - \widetilde{z}| = O(c^k)$, we can show that $|\widehat{\mathfrak{m}} - \mathfrak{m}| = O(c^k)$. Similarly, we have $|\widehat{\mathfrak{M}} - \mathfrak{M}| = O(c^{k})$. 

Combining all seven steps above, we prove the theorem.
 \hfill $\Box$

%% file: asd.tex
\section{Properties of $\mu_\rho$}  \label{sec:rho}

For a probability measure $\mu$ on $\mathbb{R}$, the Stieltjes transform of $\mu$ is defined as
\begin{equation*}
G_\mu(z) = \int_{supp(\mu)} \frac{1}{z - x} \, d\mu(x), \quad z \in \mathbb{R} \setminus supp(\mu),
\end{equation*}
where $supp(\mu)$ denotes the support of $\mu$.

The Stieltjes transform plays a crucial role in characterizing the behavior of the random measure. Under certain regularity conditions, it satisfies
$G_\mu(z) = \sum_{k=0}^{\infty} m_{k}(\mu)z^{-k}$, where $m_k(\mu) = \int x^k \, d\mu(x)$ denotes the $k$th moment of $\mu$. In general, obtaining an explicit form of the Stieltjes transform for a given $\mu$ can be challenging. However, within our framework, we have a simple analytic form for $G_{\mu_\rho}$, where $\mu_\rho$ has a density function as described in (\ref{eq:cdf}).

\begin{theorem} \label{prop:szego}
Let $\mu_\rho$ be defined as in (\ref{eq:cdf}), and let $a_\rho$ and $b_\rho$ be the lower and upper bounds of the support of $\mu_\rho$, respectively, as defined in (\ref{eq:bounds}). Then, for any $z \in [a_\rho, b_\rho]^c$, the Stieltjes transform of $\mu_\rho$ is given by
\begin{equation*}
G_{\mu_\rho}(z) = 
\begin{cases}
\frac{1}{\sqrt{(z-a_\rho)(z-b_\rho)}}, & \text{if } z > b_\rho. \\
- \frac{1}{\sqrt{(z-a_\rho)(z-b_\rho)}}, & \text{if } z < a_\rho.
\end{cases}
\end{equation*}
\end{theorem}
\noindent \textit{Proof}.
Let $\{\mu_n\}$ be a sequence of compactly supported measures on $\mathbb{R}$. It is well-known that $\mu_n \xrightarrow{D} \mu$ if and only if $G_{\mu_n}(z) \rightarrow G_{\mu}(z)$ for all $z \in \mathbb{R} \setminus \text{supp}(\mu)$. Combining this fact with Lemma \ref{lemma:LSD}, it is sufficient to show that
\begin{equation*}
\lim_{n \to \infty} G_{\mu_{\widetilde{A}_{0,n}}}(z) =
\begin{cases}
\frac{1}{\sqrt{(z - a_\rho)(z - b_\rho)}}, & \text{if } z > b_\rho, \\
- \frac{1}{\sqrt{(z - a_\rho)(z - b_\rho)}}, & \text{if } z < a_\rho,
\end{cases}
\end{equation*}
where $\widetilde{A}_{0,n}$ is defined as in (\ref{eq:perturb}).

By definition, we have
\begin{equation*}
G_{\mu_{\widetilde{A}_{0,n}}}(z) = \frac{1}{n} \sum_{i=1}^{n} \frac{1}{z - \lambda_i (\widetilde{A}_{0,n})},
\quad z \in [a_\rho, b_\rho]^c.
\end{equation*}
Let $g_{\widetilde{A}_{0,n}}(x) = (1 + \rho^2) - 2 \rho \cos x$ be the generating function of the Toeplitz matrix $\widetilde{A}_{0,n}$. Then, by applying the Szeg{\"o} limit theorem, we have
\begin{equation} \label{eq:szego.thm}
\lim_{n \to \infty} \frac{1}{n} \sum_{i=1}^{n} f(\lambda_i (\widetilde{A}_{0,n})) = \frac{1}{2\pi} \int_{0}^{2\pi} f \left( g_{\widetilde{A}_{0,n}}(x) \right) \, dx
\end{equation}
for any continuous function $f : [a_\rho, b_\rho] \rightarrow \mathbb{R}$. In particular, setting $f_z(x) = (z - x)^{-1}$ for $z \in [a_\rho, b_\rho]^c$, we have
\begin{eqnarray*}
\lim_{n \to \infty} G_{\mu_{\widetilde{A}_{0,n}}}(z) &=& \frac{1}{2\pi} \int_{0}^{2\pi} f_z\left( g_{\widetilde{A}_{0,n}}(x) \right) \, dx
= \frac{1}{2\pi} \int_{0}^{2\pi} \frac{1}{2 \rho \cos x + (z - (1 + \rho^2))} \, dx \\
&=& 
\begin{cases}
\frac{1}{\sqrt{(z - a_\rho)(z - b_\rho)}}, & \text{if } z > b_\rho. \\
- \frac{1}{\sqrt{(z - a_\rho)(z - b_\rho)}}, & \text{if } z < a_\rho.
\end{cases}
\end{eqnarray*}
Here, the last identity can be proven using similar techniques as in Lemma \ref{lemma:int} (details omitted).
\hfill $\Box$

\vspace{0.5em}

The following corollary provides an expression of the moments of $\mu_\rho$.
\begin{corollary}
\label{coro:szego}
The $k$-th moment of $\mu_\rho$ is given by
\begin{equation*}
m_k(\mu_\rho) = \frac{1}{2\pi} \int_{0}^{2\pi} (1 + \rho^2 - 2 \rho \cos x)^k \, dx, \quad k \in \mathbb{N}.
\end{equation*}
\end{corollary}
\noindent \textit{Proof}. This follows immediately from (\ref{eq:szego.thm}) by setting $f(x) = x^k$.
\hfill $\Box$


%% file: proofs.tex
\section{Additional proofs of the main results} \label{sec:proofs}
This section contains additional proofs from the main paper. For simplicity, we mainly focus on the case $\rho \in (0,1)$. The proof for $\rho \in (-1,0)$ can be treated similarly.

\subsection{Proof of Theorem \ref{prop:asymdist}} \label{prop:asymdist:proof}
We only prove the results for the case when $m=1$, and the general case when $m \geq 2$ can be treated similarly. Let $A_{0,n}$ and $B_n$ be the precision matrices under the null and SCM, respectively. We will first show (\ref{eq:LSD}) for $|E_{1,n}| = 1$, and then extend the result to the general case where $|E_{1,n}| \in \mathbb{N}$.

\vspace{0.5em}

\noindent \underline{\textbf{case 1}}: $|E_{1,n}|=1$. 

Suppose we have shown the following:
\begin{equation}
\label{eq:momentcond}
\lim_{n \rightarrow \infty} \frac{1}{n} \operatorname{tr}(B_n^j) = \lim_{n \rightarrow \infty} \frac{1}{n} \operatorname{tr}(A_{0,n}^j) = m_j(\mu_\rho), \quad j \in \{0,1,\dots\},
\end{equation}
where $m_j(\mu_\rho)$ is the $j$th moment of $\mu_\rho$. Then, by Lemma \ref{lemma:compact} together with \cite{b:bil-08}, Theorem 30.2, we get $\mu_{B_n} \Dcon \mu_{\rho}$ as $n \rightarrow \infty$. Therefore, it suffices to show (\ref{eq:momentcond}).

Let $R_n = B_n - A_{0,n}$. Then, from (\ref{eq:Anform}), the entries of $R_n$ are zero except for a $2 \times 2$ submatrix. Next, by using linearity, we have
\begin{equation}
\operatorname{tr}(B_n^j) = \operatorname{tr} \left( (A_{0,n} + R_n)^j \right) = \sum_{\alpha_i \in \{\circ,~\ast\} } \operatorname{tr} \left( X_n^{(\alpha_1)} \cdots X_n^{(\alpha_j)} \right),
\end{equation} 
where $X_n^{(\circ)} = A_{0,n}$ and $X_n^{(\ast)} = R_n$. 

Observe that $R_n$ has nonzero elements in $[R_n]_{i,j}$ for $(i,j) \in \{(k-1,k-1),(k-1,k),(k,k-1)\}$, where $k = k_n$ is the breakpoint. Therefore, for any matrix $X \in \mathbb{R}^{n \times n}$, the columns of $XR_n$ are zero vectors except for the $(k-1)$th and $k$th columns. Next, using the commutativity of the trace operator, we get
\begin{equation} \label{eq:trX}
\operatorname{tr} \left(X_n^{(\alpha_1)} \cdots X_n^{(\alpha_j)} \right) = \operatorname{tr} \left( A_{0,n}^{n_1} R_n^{m_1} \cdots A_{0,n}^{n_t} R_n^{m_t} \right),
\end{equation} 
for some orders $(n_1, \dots, n_t, m_1, \dots, m_t)$. Since $A_{0,n}^{n_p} R_n^{m_p}$ has at most two nonzero columns (specifically, the $(k-1)$th and $k$th columns), and so does the product. Therefore, $ X_n^{(\alpha_1)} \cdots X_n^{(\alpha_j)}$ has at most two nonzero diagonal elements unless all $X_n^{(\alpha_i)}$ are $A_{0,n}$. Hence, we can find a constant $B_j \in (0,\infty)$ that does not depend on $n$ such that 
\begin{equation*}
\max_{1 \leq i \leq n} \left| \left[ X_n^{(\alpha_1)} \cdots X_n^{(\alpha_j)} \right]_{i,i} \right| < B_j, \quad \text{for all $\{\alpha_i\}$ that are not all equal to $\circ$}.
\end{equation*} 
Using this bound, we have
\begin{equation*}
\lim_{n \rightarrow \infty} \frac{1}{n} \left| \operatorname{tr}(B_n^j) - \operatorname{tr}(A_{0,n}^j) \right| \leq \lim_{n \rightarrow \infty} (2^j - 1) \times \frac{2B_j}{n} = 0, \quad j \in \{0,1, \dots\}.
\end{equation*}
This proves (\ref{eq:LSD}) for $|E_{1,n}| = 1$.

\vspace{0.5em}

\noindent \underline{\textbf{case 2}}: $|E_{1,n}|>1$ and $\lim_{n\rightarrow \infty} |E_{1,n}|/n = \tau = 0$.

By using a similar method as above, there exists $\widetilde{B}_j \in (0,\infty)$ that does not depend on $n$ such that 
\begin{equation*}
\frac{1}{n} \left| \operatorname{tr}(B_n^j) - \operatorname{tr}(A_{0,n}^j) \right| \leq \frac{2^j}{n} (|E_{1,n}| + 1) \widetilde{B}_j, \quad j \in \{0,1, \dots\}.
\end{equation*}
Since $\lim_{n \rightarrow \infty} |E_{1,n}|/n = \tau = 0$, the right-hand side above converges to zero, thus proving (\ref{eq:LSD}).

\vspace{0.5em}

\noindent \underline{\textbf{case 3}}: $|E_{1,n}|>1$ and $\tau \in (0,1)$.

Let the structural change occur on $\{k, \dots, k+h-1\}$ for some $h \in \{2, 3, \dots\}$. Define the $n \times n$ matrix
\begin{equation}
(P_n)_{i,j}=
\left\{
\begin{array}{ll}
-\rho_{k-1}^2, & \text{if } (i,j) = (k-2,k-2). \\
\rho_{k}, & \text{if } (i,j) = (k-1,k), (k,k-1). \\
-\rho_{k+h-1}^2, & \text{if } (i,j) = (k+h-2,k+h-2). \\
\rho_{k+h}, & \text{if } (i,j) = (k+h-1,k+h), (k+h,k+h-1). \\
0, & \text{otherwise}.
\end{array}
\right.
\end{equation}
Then, $\operatorname{rank}(P_n) \leq 4$, so $P_n$ has at most four nonzero eigenvalues.

Let $\widetilde{B}_n = B_n + P_n$. It is easily seen that $\widetilde{B}_n$ is a block diagonal matrix of the form
\begin{equation*}
\widetilde{B}_n = \diag (\widetilde{B}_{1,n}, \widetilde{B}_{2,n}, \widetilde{B}_{3,n}),
\end{equation*}
where $\widetilde{B}_{i,n}$ forms an inverse Toeplitz matrix of the null model but with different baseline AR coefficients. Specifically, $\widetilde{B}_{1,n}$ and $\widetilde{B}_{3,n}$ correspond to the precision matrices of the null model with the baseline AR coefficient $\rho$, and $\widetilde{B}_{2,n}$ corresponds to the null model with the baseline AR coefficient $\rho + \varepsilon$.

Since $P_n$ has a finite number of nonzero eigenvalues, by using the same proof techniques as above, the LSD of $B_n$ and $\widetilde{B}_n$ are the same. Moreover, since $\widetilde{B}_n$ is a block diagonal matrix, we have for all $j \in \{0,1, \dots\}$,
\begin{equation*}
\lim_{n \rightarrow \infty} \frac{1}{n} \operatorname{tr}(\widetilde{B}_n^j) = \lim_{n \rightarrow \infty} \frac{1}{n} \left( \operatorname{tr}(\widetilde{B}_{1,n}^j) 
+ \operatorname{tr}(\widetilde{B}_{2,n}^j) + \operatorname{tr}(\widetilde{B}_{3,n}^j) \right) 
= \tau m_{j}(\mu_{\rho+\varepsilon}) + (1-\tau) m_{j}(\mu_{\rho}).
\end{equation*}
Thus, we obtain the desired results.
\hfill $\Box$

\subsection{Proof of Theorem \ref{thm:multi_determinant}} \label{thm:multi_determinant:proof}

We only prove the theorem for the case $\rho \in (0,1)$, and the case when $\rho \in (-1,0)$ can be treated similarly. The proof of the theorem is similar to that of Theorem \ref{thm:spacing}, so we only sketch the proof, which is divided into two steps.

\subsubsection*{Step 1}
Let $A_{0,n}$ and $B_n$ be the precision matrices under the null and single interval SCM, respectively. Let $\widetilde{A}_{0,n}$ be defined as in (\ref{eq:perturb}), and let $\widetilde{A}_{0,n} := U_n \Lambda_n U_n^\top$ be its eigen-decomposition as described in Section \ref{thm:spacing:proof}, [Step 2]. Let $M_{n,r}$ be as in (\ref{eq:M_nr}). Then, according to Section \ref{thm:spacing:proof}, [Step 2], $z$ is an eigenvalue of $B_n$ but not of $\widetilde{A}_{0,n}$ if and only if $M_{n,r}$ is singular. In the single interval SCM, the rank $r = h + 2$, where $h \in \mathbb{N}$ is the length of the change. Moreover, we have the following reduced form $P_{n} = B_n - A_{0,n}$ by considering only the nonzero submatrix in (\ref{eq:pform}):
\begin{equation}
\label{eq:predform}
P_{h+1} = \varepsilon \begin{pmatrix}
\varepsilon + 2\rho & -1 & & & \\ 
-1 & \varepsilon + 2\rho & \ddots & & \\
& \ddots & \ddots & & \\
& & & \varepsilon + 2\rho & -1 \\
& & & -1 & 0
\end{pmatrix} \in \mathbb{R}^{(h+1) \times (h+1)}.
\end{equation}

Let $P_{h+1} = V_{h+1} \Theta_{h+1} V_{h+1}^\top$ be the spectral decomposition. Then, by similar arguments from Section \ref{thm:spacing:proof}, [Step 3], we have the $(h+1) \times (h+1)$ leading principal matrix of $M_{n,r}$, denoted $M_{n,h+1}$, given by $M_{n,h+1} = I_{h+1} - S_{h+1}^\top (zI_n - \Lambda_n)^{-1} S_{h+1} \Theta_{h+1}$. Here, $S_{h+1} = ( s_1, \ldots, s_{h+1})$ for $s_i = \sum_{j=1}^{h+1} v_{j,i} u_{k+j}$. Here, $U_n = ( u_1, \ldots, u_{n})$ is as in Section \ref{thm:spacing:proof}, [Step 2], and $v_{j,i}$ denotes the $(j,i)$th element of $V_{h+1} = ( v_1, \ldots, v_{h+1} )$.

Next, define $M_{h+1} = \lim_{n \rightarrow \infty} M_{n,h+1}$, then the possible outliers of $B_n$ are the solutions to $\det M_{h+1} = 0$. By (\ref{eq:Mmatrix}), the $(i,j)$th element of $M_{h+1}$ is given by
\begin{equation*}
[M_{h+1}]_{i,j} = \delta_{i=j} - \theta_j \lim_{n \rightarrow \infty} \sum_{\ell = 1}^{n} \frac{[S_n]_{\ell, i} [S_n]_{\ell, j} }{2\rho \cos\left( \frac{\ell \pi}{n+1} \right)+ \left( z - (1+\rho^2) \right) }.
\end{equation*}
Observe that
\begin{eqnarray*}
\lim_{n \rightarrow \infty}\sum_{\ell = 1}^{n} \frac{[S_n]_{\ell, i} [S_n]_{\ell, j}}{\cos \left(\frac{\ell \pi}{n+1}\right) + a } &=& 
\sum_{p,q = 1}^{h+1} v_{p,i} v_{q,j} \lim_{n \rightarrow \infty} \sum_{\ell = 1}^{n} \frac{2}{n+1} \frac{\sin \left(\frac{ (k+p) \ell \pi}{n+1}\right) \sin \left(\frac {(k+q) \ell \pi}{n+1} \right)} {\cos \left( \frac{\ell \pi}{n+1} \right)+ a}\\
&=& \sum_{p,q = 1}^{h+1} v_{p,i} v_{q,j} G(k+p,k+q) = v_i^{\top} G_{h+1} v_j,
\end{eqnarray*} 
where $G_{h+1} = [G(k+i, k+j)]_{i,j} \in \mathbb{R}^{(h+1) \times (h+1)}$ and $G(k+i,k+j)$ is defined as in (\ref{eq:sineint1}). 
Therefore, the possible outliers of $B_n$ satisfy the determinantal equation
\begin{equation} \label{eq:vecM}
\det \left( I_{h+1} - \frac{1}{2\rho} V_{h+1}^\top G_{h+1} V_{h+1} \Theta_{h+1} \right) = 0.
\end{equation}

\subsubsection*{Step 2}

Since $V_{h+1}^\top V_{h+1} = V_{h+1} V_{h+1}^\top = I_{h+1}$ and $V_{h+1} \Theta_{h+1} V_{h+1}^\top = P_{h+1}$, solving (\ref{eq:vecM}) is equivalent to solving
\begin{equation} \label{eq:GP}
\det \left( I_{h+1} - \frac{1}{2\rho} G_{h+1} P_{h+1} \right) = 0.
\end{equation}
For $z > b_\rho$ (the case when $z < a_\rho$ can be treated similarly), by Lemma \ref{lemma:int}, the explicit form of an element of $G_{h+1}$ is given by
\begin{equation*}
\frac{1}{2} [G_{h+1}]_{p,q} = \frac{1}{2} G(k+p,k+q) = z_2 (z_2^2 - 1)^{-1} (z_2^{|p-q|} - z_2^{p+q+2k}) = 
z_2 (z_2^2 - 1)^{-1} z_2^{|p-q|} + O\left( \frac{|z_2|^k}{z_2^2 - 1} \right).
\end{equation*}
We note that the error in the right-hand side above vanishes, provided (\ref{assum:SCM}).

Next, we observe that the leading term $\frac{1}{2} [G_{h+1}]_{p,q} = z_2 (z_2^2 - 1)^{-1} z_2^{|p-q|}$ has the same form (up to a constant multiplicity) as the covariance matrix of a stationary AR$(1)$ process. Therefore, an explicit form of its inverse is
\begin{equation}\label{eq:Ginv}
\left( \frac{1}{2} G_{h+1} \right)^{-1} = -\frac{1}{z_2} \begin{pmatrix}
1 & -z_2 & & \\
-z_2 & 1 + z_2^2 & -z_2 & \\
& -z_2 & \ddots & \ddots \\
& & \ddots & 1 + z_2^2 & -z_2 \\
& & & -z_2 & 1
\end{pmatrix}.
\end{equation}
Since $\det \left(- \frac{1}{2} G_{h+1} \right) \neq 0$, solving (\ref{eq:GP}) is equivalent to solving $\det \left(-\left(\frac{1}{2} G_{h+1}\right)^{-1} + \frac{1}{\rho} P_{h+1} \right) = 0$. Using (\ref{eq:Ginv}), we get
\begin{equation} \label{eq:sparse}
-\left( \frac{1}{2} G_{h+1} \right)^{-1} + \frac{1}{\rho} P_{h+1} 
= \left(1+ \frac{\varepsilon}{\rho}\right) \begin{pmatrix}
\alpha & -1 & & \\
-1 & \beta & \ddots & \\
& \ddots & \ddots & \ddots \\
& & \ddots & \beta & -1 \\
& & & -1 & \gamma
\end{pmatrix} \in \mathbb{R}^{(h+1) \times (h+1)},
\end{equation}
where $\alpha$, $\beta$, and $\gamma$ are defined as in (\ref{eq:param}).

Lastly, note that the actual outlier is $z = 1 + \rho^2 - \rho \left( z_2 + z_2^{-1} \right)$. It is easy to check that $z \notin [a_\rho, b_\rho]$ if and only if (\ref{eq:sparse}) holds for $f^{-1}(z) \in (-1,1)$, where $f$ is as defined in (\ref{eq:ftrans}). Thus, we get the desired results. \hfill $\Box$

\subsection{Proof of Theorem \ref{thm:block}} \label{thm:block:proof}
For $j \in \{1, \dots, m\}$, let $P_{h_j+1}^{(j)} \in \mathbb{R}^{(h_{j}+1) \times (h_{j}+1)}$ be defined as in (\ref{eq:predform}), but with $\varepsilon$ replaced by $\varepsilon_{j}$. Let $0_{r} \in \mathbb{R}^{r \times r}$ be the zero matrix and let
\begin{equation*}
P_{n} = \diag\left(0_{\Delta_{1}-2}, P_{h_1+1}^{(1)}, \dots, 0_{\Delta_{m}-2}, P_{h_m+1}^{(m)}, 0_{n-\ell_{m}}\right) \in \mathbb{R}^{n \times n}.
\end{equation*}
Then, it is easy to show that $P_{n}=B_{n}-A_{0,n}$, where $A_{0,n}$ is defined as in (\ref{eq:An.null}). Let
$P_{h+m} = \diag\big( P_{h_1+1}^{(1)}, \dots, P_{h_m+1}^{(m)} \big) \in \mathbb{R}^{(h+m) \times (h+m)}$,
where $h = \sum_{j=1}^{m} h_{j}$, be the reduced form of $P_{n}$.

Given $i \in \{1, \dots, h+m\}$, let $\ell(i) \in \{1, \dots, m\}$ be the unique index such that 
\begin{equation*}
\sum_{a=1}^{\ell(i)-1}(h_{a}+1) < i \leq \sum_{a=1}^{\ell(i)} (h_{a}+1).
\end{equation*}
Here, we set $\sum_{a=1}^{0}(h_{a}+1) = 0$. Let $g(i) = h_{\ell(i)} + \big( i-\sum_{a=1}^{\ell(i)-1}(h_{a}+1)\big)$. Then, $g(i)$ is the location of the column of $P_{n}$ which is the same as the $i$th column of $P_{R}$.

Next, similar to the proof of Theorem \ref{thm:multi_determinant}, [Step 1], we can show that the corresponding $G_{h+m} \in \mathbb{R}^{(h+m) \times (h+m)}$ matrix of $P_{R}$ is
\begin{equation*}
[G_{h+m}]_{i,j} = G(g(i), g(j)), \quad 1 \leq i, j \leq h+m,
\end{equation*}
where $G(\cdot, \cdot)$ is defined as in (\ref{eq:sineint1}). Therefore, using similar arguments to those in the proof of Theorem \ref{thm:multi_determinant}, [Step 2], there exists $c \in (0,1)$ such that
\begin{equation*}
\frac{1}{2}[G_{h+m}]_{i,j} = \left \{
\begin{array}{ll}
z_{2} (z_{2}^{2}-1)^{-1} z_{2}^{|i-j|}, & \text{if } \ell(i)=\ell(j), \\
0, & \text{if } \ell(i)\neq \ell(j)
\end{array}
\right. + O(c^{\Delta}), \quad n \rightarrow \infty.
\end{equation*}
Therefore, under assumption \ref{assum:SCM2}, the leading term of $G_{h+m}$ is a block diagonal matrix of the form $\diag(G_{h_1+1}^{(1)}, \ldots, G_{h_m+1}^{(m)})$, where $G_{h_j+1}^{(j)} \in \mathbb{R}^{(h_{j}+1) \times (h_{j}+1)}$ corresponds to the $G$ matrix of the $j$th submodel defined as in the proof of Theorem \ref{thm:multi_determinant}, [Step 1].


Next, by using simiar techniques to those in the proof of Theorem \ref{thm:multi_determinant}, [Step 2], outliers of $B_{n}$ are the zeros of the determinantal equation
\begin{equation}\label{eq:Ghm} 
\det \left( I_{h+m} - \frac{1}{2\rho} G_{h+m} P_{h+m} \right) = 0.
\end{equation} Since $G_{h+m}$ and $P_{h+m}$
 are block diagonal matrix, (\ref{eq:Ghm}) is equivalant to solve
\begin{equation*}
\prod_{j=1}^{m}\det \left( I_{h_j+1} - \frac{1}{2\rho} G_{h_j+1}^{(j)} P_{h_j+1} \right) = 0.
\end{equation*}

Lastly, from the proof of Theorem \ref{thm:multi_determinant}, the zeros of $\det \left( I_{h_j+1} - \frac{1}{2\rho} G_{h_j+1}^{(j)} P_{h_j+1} \right) = 0$ are indeed the outliers of the $j$th submodel. Thus, we have
\begin{equation*}
\text{out}(\{B_n\}) = \bigcup_{j=1}^{m} \text{out}(\{B_n^{(j)}\}),
\end{equation*}
which proves the theorem.
\hfill $\Box$

\subsection{Proof of Theorem \ref{thm:consist}} \label{thm:consist:proof}
For a set $A \subset \mathbb{R}$, let
\begin{equation*}
out_{L}(A) = A \cup (-\infty, a_\rho), \quad
out_{R}(A) = A \cup (b_\rho, \infty), \quad \text{and} \quad
out(A) = A \cup [a_\rho, b_\rho]^{c}.
\end{equation*}
We define $\widehat{out}_{L}(A)$ and $\widehat{out}_{R}(A)$ similarly, but replacing $a_\rho$ and $b_\rho$ with $a_{\widehat{\rho}_n}$
and $b_{\widehat{\rho}_n}$, respectively.

By the triangle inequality, we have
\begin{equation}
\begin{aligned} 
d_{H}\left( \widehat{out}(\widetilde{\Omega}_{n,B}), \widehat{out}(\{\Omega_{n}\}) \right) &\leq
d_{H}\left( \widehat{out}(\widetilde{\Omega}_{n,B}), \widehat{out}(\Omega_{n}) \right) + 
d_{H}\left( \widehat{out}(\Omega_{n}), out(\Omega_{n}) \right) \\
&\quad + d_{H}\left( out(\Omega_{n}), out(\{\Omega_{n}\}) \right).
\end{aligned}
\label{eq:tria}
\end{equation}
Now, we bound the three terms above. The last term in (\ref{eq:tria}) is non-random, and by the definition of outliers,
\begin{equation}\label{eq:tribound1}
d_{H}\left( out(\Omega_{n}), out(\{\Omega_{n}\}) \right) \rightarrow 0, \quad n \rightarrow \infty.
\end{equation}

Next, we bound the second term in (\ref{eq:tria}). Here, we only consider the case when $out_L(\{\Omega_{n}\})$ and $out_R(\{\Omega_{n}\})$ are nonempty. The case when either set is empty is straightforward. Let $a = \sup out_L(\{\Omega_n\})$, thus $a < a_\rho$. Let $\eta = (a_{\rho} - a)/2 \in (0, \infty)$ and let $\delta > 0$. Then,
\begin{eqnarray*}
P\left( d_{H}\left( \widehat{out}_{L}(\Omega_n), out_{L}(\Omega_{n}) \right) > \delta \right) &=& 
P\left( d_{H}\left( \widehat{out}_{L}(\Omega_n), out_{L}(\Omega_{n}) \right) > \delta \mid |a_{\widehat{\rho}_n} - a_{\rho}| > \eta \right) \\
&& \times
P(|a_{\widehat{\rho}_n} - a_{\rho}| > \eta) \\
&+& 
P\left( d_{H}\left( \widehat{out}_{L}(\Omega_n), out_{L}(\Omega_{n}) \right) > \delta \mid |a_{\widehat{\rho}_n} - a_{\rho}| \leq \eta \right) \\
&& \times
P(|a_{\widehat{\rho}_n} - a_{\rho}| \leq \eta).
\end{eqnarray*}

If $|a_{\widehat{\rho}_n} - a_{\rho}| \leq \eta$, then for large enough $n \in \mathbb{N}$,
$\sup out_L(\Omega_n) < a_{\widehat{\rho}_n}$. Thus, $out_L(\Omega_n) = \widehat{out}_L(\Omega_n)$
and $d_{H}\left( \widehat{out}_{L}(\Omega_n), out_{L}(\Omega_{n}) \right) = 0$. Therefore,
for large enough $n \in \mathbb{N}$,
\begin{eqnarray*}
P\left( d_{H}\left( \widehat{out}_{L}(\Omega_n), out_{L}(\Omega_{n}) \right) > \delta \right) &=& 
P\left( d_{H}\left( \widehat{out}_{L}(\Omega_n), out_{L}(\Omega_{n}) \right) > \delta \mid |a_{\widehat{\rho}_n} - a_{\rho}| > \eta \right) \\
&& \times
P(|a_{\widehat{\rho}_n} - a_{\rho}| > \eta) \\
&\leq& P(|a_{\widehat{\rho}_n} - a_{\rho}| > \eta).
\end{eqnarray*}
Now, by using the continuous mapping theorem, $P(|a_{\widehat{\rho}_n} - a_{\rho}| > \eta) \rightarrow 0$ as $n \rightarrow \infty$. Thus, we have
$d_{H}\big( \widehat{out}_{L}(\Omega_n), out_{L}(\Omega_{n}) \big) \Pcon 0$. Similarly, we can show
$d_{H}\big( \widehat{out}_{R}(\Omega_n), out_{R}(\Omega_{n}) \big) \Pcon 0$. Since the left and right
outliers are disjoint, we have
\begin{equation*}
d_{H}\left( \widehat{out}(\Omega_{n}), out(\Omega_{n}) \right) = 
d_{H}\left( \widehat{out}_{L}(\Omega_n), out_{L}(\Omega_{n}) \right) \vee
d_{H}\left( \widehat{out}_{R}(\Omega_n), out_{R}(\Omega_{n}) \right).
\end{equation*}
Therefore, we conclude
\begin{equation} \label{eq:tribound2}
d_{H}\left( \widehat{out}(\Omega_{n}), out(\Omega_{n}) \right) \Pcon 0, \quad n \rightarrow \infty.
\end{equation}

Lastly, we bound the first term in (\ref{eq:tria}). Let $\delta \in (0,\infty)$. Then,
\begin{equation*}
P\left( d_{H}\left( \widehat{out}(\widetilde{\Omega}_{n,B}), \widehat{out}(\Omega_{n}) \right) \leq \delta \right) \geq 
P\left( d_{H}\left( \widehat{out}(\widetilde{\Omega}_{n,B}), \widehat{out}(\Omega_{n}) \right) \leq \delta, |\widehat{out}(\widetilde{\Omega}_{n,B})| = |\widehat{out}(\Omega_{n})| \right).
\end{equation*} 
By using Lemma \ref{lemma:consist}, it can be shown that for large enough $n \in \mathbb{N}$, $|\widehat{out}(\widetilde{\Omega}_{n,B(n)})| = |\widehat{out}(\Omega_{n})|$ with probability greater than $(1 - 4n^{-1/2})$. Therefore, for large enough $n \in \mathbb{N}$ and given that $|\widehat{out}(\widetilde{\Omega}_{n,B(n)})| = |\widehat{out}(\Omega_{n})| = \ell$,
\begin{eqnarray*}
d_{H}\left( \widehat{out}(\widetilde{\Omega}_{n,B(n)}), \widehat{out}(\Omega_{n}) \right) &=&
\max_{1 \leq i \leq \ell} |\lambda_{t_{i}}(\widetilde{\Omega}_{n,B(n)}) - \lambda_{t_{i}}(\Omega_{n}) | \\
&\leq& \max_{1 \leq i \leq n} |\lambda_{i}(\widetilde{\Omega}_{n,B(n)}) - \lambda_{i}(\Omega_{n}) |,
\end{eqnarray*}
where $t_{1}, \ldots, t_{\ell}$ are indices of eigenvalues which are outliers. 
This indicates that
\begin{eqnarray*}
&&P\left( d_{H}\left( \widehat{out}(\widetilde{\Omega}_{n,B(n)}), \widehat{out}(\Omega_{n}) \right) \leq \delta, |\widehat{out}(\widetilde{\Omega}_{n,B(n)})| = |\widehat{out}(\Omega_{n})| \right) \\
&&\quad \geq
P\left(\max_{1 \leq i \leq n} |\lambda_i (\widetilde{\Omega}_{n,B(n)}) - \lambda_i(\Omega_{n})| \leq \delta, |\widehat{out}(\widetilde{\Omega}_{n,B(n)})| = |\widehat{out}(\Omega_{n})| \right).
\end{eqnarray*}
Next, by using Lemma \ref{lemma:consist} again, for large enough $n$,
\begin{eqnarray*}
&&P\left( d_{H}\left( \widehat{out}(\widetilde{\Omega}_{n,B(n)}), \widehat{out}(\Omega_{n}) \right) \leq \delta \right) \\
&&\quad \geq
P\left(\max_{1 \leq i \leq n} |\lambda_i (\widetilde{\Omega}_{n,B(n)}) - \lambda_i(\Omega_{n})| \leq \delta, |\widehat{out}(\widetilde{\Omega}_{n,B(n)})| = |\widehat{out}(\Omega_{n})| \right) > 1 - 8n^{-1/2}.
\end{eqnarray*}
This implies
\begin{equation} \label{eq:tribound3}
d_{H}\left( \widehat{out}(\widetilde{\Omega}_{n,B(n)}), \widehat{out}(\Omega_{n}) \right) \Pcon 0, \quad n \rightarrow \infty.
\end{equation}

Combining (\ref{eq:tribound1}), (\ref{eq:tribound2}), and (\ref{eq:tribound3}), we get
\begin{equation*}
d_{H}\left( \widehat{out}(\widetilde{\Omega}_{n,B}), out(\{\Omega_{n}\}) \right) \Pcon 0, \quad n \rightarrow \infty,
\end{equation*}
which proves the theorem. \hfill $\Box$

%% file: technical_detail.tex
\section{Technical Lemmas} \label{sec:tech}

\begin{lemma}
\label{lemma:int}
Let $a \in \mathbb{R}$ be a constant such that $|a| > 1$ and let
\begin{equation} \label{eq:z12}
z_1 = -a - \sqrt{a^2-1} 
\quad \text{and} \quad
z_2 = -a + \sqrt{a^2 - 1}.
\end{equation} 
Then, for any $k_1, k_2 \in \N$, we have
\begin{equation}
\label{eq:sineint1}
\frac{1}{2} G(k_1, k_2) := \frac{1}{2\pi} \int_{0}^{2 \pi} \frac{\sin (k_1 x) \sin (k_2 x)}{a + \cos x} dx = \left 
\{\begin{array}{ll}
\frac{1}{z_2 - z_1} \big( z_2^{|k_1 - k_2|}-    z_2^{k_1+k_2} \big), & a>1. \\
\frac{1}{z_1 - z_2} \big( z_1^{|k_1 - k_2|}-    z_1^{k_1+k_2} \big), & a<-1. \end{array}
  \right. 
\end{equation}
Therefore, for fixed $h \in \mathbb{N}$, as $k \rightarrow \infty$,
\begin{equation*}
G(k,k+h) = 
\left 
\{\begin{array}{ll}
\frac{a}{|a|} \frac{1}{ \sqrt{a^2-1}} z_2^h + O(|z_{2}|^{k}), & a>1 \\
\frac{a}{|a|} \frac{1}{ \sqrt{a^2-1}} z_1^h + O(|z_{1}|^{k}), & a<-1
\end{array}
  \right.
\end{equation*}
and $\lim_{k,h \rightarrow \infty} G(k, k+h) = 0$.
\end{lemma}

\noindent \textit{Proof}. 
We will only prove the identities for $a > 1$, and the case when $a < -1$ can be treated similarly. Let $z = e^{ix}$ where $i = \sqrt{-1}$. Then, we have (1) $dz = iz dx$, (2) $\cos x = \frac{1}{2} (z + z^{-1})$, and (3) $\sin(kx) = \frac{1}{2i} (z^k - z^{-k})$. Let $C$ be a counterclockwise contour of the unit circle in the complex plane starting from 1, and let $\oint_{C}$ denote a contour integral along $C$. Then,
\begin{eqnarray*}
\frac{1}{2\pi} \int_{0}^{2\pi} \frac{\sin(k_1 x) \sin(k_2 x)}{a + \cos x} \, dx &=& \frac{1}{2\pi} \oint_C \left( \frac{-\frac{1}{4} (z^{k_1} - z^{-k_1}) (z^{k_2} - z^{-k_2})}{\frac{1}{2} (z + z^{-1}) + a} \right) \frac{dz}{iz} \\
&=& -\frac{1}{2} \frac{1}{2\pi i} \oint_C \frac{(z^{k_1} - z^{-k_1})(z^{k_2} - z^{-k_2})}{(z - z_1)(z - z_2)} \, dz.
\end{eqnarray*}

Since $a > 1$, we have $|z_2| < 1 < |z_1|$. Therefore, the poles of $(z^{2k_1}-1) (z^{2 k_2}-1) / (z^{k_1+k_2} (z-z_1)(z-z_2))$ inside the contour $C$ are $z_2$ with multiplicity 1, and $0$ with multiplicity $(k_1 + k_2)$. Therefore, by using Cauchy's integral formula,
\begin{eqnarray*}
&& \frac{1}{2\pi i} \oint_C \frac{(z^{2k_1} - 1)(z^{2k_2} - 1)}{z^{k_1 + k_2}(z - z_1)(z - z_2)} \, dz \\
&&= \text{Res} \left( \frac{(z^{2k_1} - 1)(z^{2k_2} - 1)}{z^{k_1 + k_2}(z - z_1)(z - z_2)}, z_2 \right) 
+ \frac{1}{(k_1 + k_2 - 1)!} \left. \frac{d^{k_1 + k_2 - 1}}{dz^{k_1 + k_2 - 1}} \left( \frac{(z^{2k_1} - 1)(z^{2k_2} - 1)}{z^{k_1 + k_2}(z - z_1)} \right) \right|_{z = 0} \\
&&= \frac{(z_2^{k_1} - z_1^{k_1})(z_2^{k_2} - z_1^{k_2})}{z_2 - z_1} 
+ \frac{1}{(k_1 + k_2 - 1)!} \left. \frac{d^{k_1 + k_2 - 1}}{dz^{k_1 + k_2 - 1}} \left( \frac{(z^{2k_1} - 1)(z^{2k_2} - 1)}{z^{k_1 + k_2}(z - z_1)} \right) \right|_{z = 0}.
\end{eqnarray*}
Here, $\text{Res}$ denotes the residue, $a(z) =  (z^{2k_1}-1) (z^{2 k_2}-1) / \{(z-z_1)(z-z_2)\}$, and $a^{(n)}$ denotes the $n$th derivative of $a(\cdot)$. 
For the second equality, we use the fact that $z_2^{-1} = z_1$. Next, since $|z/z_1|, |z/z_2| < 1$ for $z$ near the origin, we have the following Taylor expansion of $a(z)$ at $z=0$:
\begin{equation}
\begin{aligned}
a(z) &= \frac{1}{z_2 - z_1} \left( z^{2(k_1 + k_2)} - z^{2k_1} - z^{2k_2} + 1 \right) 
\left[ \frac{1}{z_1} \left( \frac{1}{1 - z/z_1} \right) - \frac{1}{z_2} \left( \frac{1}{1 - z/z_2} \right) \right] \\
&= \frac{1}{z_2 - z_1} \left( z^{2(k_1 + k_2)} - z^{2k_1} - z^{2k_2} + 1 \right) 
\left[ \frac{1}{z_1} \sum_{j=0}^{\infty} \left( \frac{z}{z_1} \right)^j - \frac{1}{z_2} \sum_{j=0}^{\infty} \left( \frac{z}{z_2} \right)^j \right].
\end{aligned}
\label{eq:taylor}
\end{equation}

Next, without loss of generality, assume $k_1 \leq k_2$. Noting that $\frac{1}{(k_1 + k_2 - 1)!} a^{(k_1 + k_2 - 1)}(0)$ is the coefficient of $z^{k_1 + k_2 - 1}$ in the power series expansion of $a(z)$ at $z=0$, we have the following two cases: (a) $k_1=k_2 = k$ and (b) $k_1 < k_2$.

For the first case, we have $1 \leq k_1 + k_2 - 1 < \{2(k_1 + k_2), 2k_1, 2k_2\}$, thus the coefficient of $z^{k_1 + k_2 -1}$ in (\ref{eq:taylor}) is
\begin{equation*}
\frac{1}{(k_1+k_2-1)!} a^{(k_1 + k_2 -1)}(0) = \frac{1}{z_2 - z_1} \left( z_1^{-2k} - z_2^{-2k} \right) = \frac{1}{z_2 - z_1} (z_2^{2k} - z_1^{2k}).
\end{equation*}
Therefore,
\begin{eqnarray*}
\frac{1}{2\pi} \int_{0}^{2 \pi} \frac{\sin(k_1 x) \sin(k_2 x)}{a + \cos x} dx &=&  -\frac{1}{2} \frac{1}{2\pi i} \oint_C \frac{(z^{2k_1}-1) (z^{2 k_2}-1)   }    { z^{k_1+k_2} (z-z_1)(z-z_2)} dz \\
&=& -\frac{1}{2} \frac{1}{(z_2 - z_1)} \big[ (z_2^{k} - z_1^{k})^2 + (z_2^{2k} - z_1^{2k} ) \big] = \frac{1}{z_2 - z_1} \left( 1-    z_2^{2k} \right).
\end{eqnarray*} 

For the second case, we have $\{1, 2k_1\} \leq k_1 + k_2 - 1 < \{2(k_1 + k_2), 2k_2\}$, thus the coefficient of $z^{k_1 + k_2 -1}$ in (\ref{eq:taylor}) is
\begin{eqnarray*}
\frac{1}{(k_1+k_2-1)!} a^{(k_1 + k_2 -1)}(0) &=& \frac{1}{z_2 - z_1} \left( z_1^{-(k_1 + k_2)} - z_1^{-(k_2 - k_1)} - 
z_2^{-(k_1+k_2)} + z_2^{-(k_2 - k_1)} \right) \\
&=& \frac{1}{z_2 - z_1} (z_2^{k_1+k_2} - z_1^{k_1} z_2^{k_2} - z_1^{k_1+k_2} + z_1^{k_2}z_2^{k_1}).
\end{eqnarray*}
Therefore,
\begin{eqnarray*}
\frac{1}{2\pi} \int_{0}^{2 \pi} \frac{\sin(k_1 x) \sin(k_2 x)}{a + \cos x} dx &=&   -\frac{1}{2} \frac{1}{(z_2 - z_1)} \bigg[ (z_2^{k_1} - z_1^{k_1}) (z_2^{k_2} - z_1^{k_2}) + \\
&& (z_2^{k_1+k_2} - z_1^{k_1} z_2^{k_2} - z_1^{k_1+k_2} + z_1^{k_2}z_2^{k_1}) \bigg] \\
&=&  \frac{1}{z_2 - z_1} \left( z_2^{k_2 - k_1}-    z_2^{k_1 + k_2} \right).
\end{eqnarray*} 
In both cases,  
\begin{equation*}
\frac{1}{2\pi} \int_{0}^{2 \pi} \frac{\sin(k_1 x) \sin(k_2 x)}{a + \cos x} dx =  \frac{1}{z_2 - z_1} \left( z_2^{|k_1 - k_2|}-    z_2^{k_1 + k_2} \right),
\end{equation*}
which proves the lemma. \hfill $\Box$

\begin{lemma} \label{lemma:eigenvalues}
Let $(a_{1}, b_{1})^{\top}$ and $(a_{2}, b_{2})^{\top}$ be the two orthonormal eigenvectors of the matrix
\begin{equation*}
\varepsilon \begin{pmatrix}
\varepsilon + 2\rho & -1 \\
-1 & 0 \\
\end{pmatrix},
\end{equation*}
where the corresponding eigenvalues are $\theta_1$ and $\theta_2$. Then, the following hold:
\begin{equation*}
\begin{aligned}
& (a)~ a_1^2 + b_1^2 = a_2^2 + b_2^2 = 1, ~~ (b)~ a_1a_2+b_1b_2 = 0, ~~ (c)~ a_1b_2 - a_2b_1 = \pm 1, \\
& (d)~ a_1^2 = b_2^2, ~~ (e)~ a_2^2 = b_1^2, ~~ (f)~ \theta_1 a_1 b_1 + \theta_2 a_2 b_2 = -\varepsilon (a_1^2+a_2^2) = -\varepsilon,
~~ (g)~ \theta_1 \theta_2 = -\varepsilon^2.
\end{aligned}
\end{equation*}
\end{lemma}
\noindent \textit{Proof}. The proof is elementary. We omit the details. \hfill $\Box$

\begin{lemma}[Weyl inequalities] \label{lemma:weyl}
Let $A_n$ and $B_n$ be $n \times n$ Hermitian matrices, and let $X_n = A_n - B_n$. Let $\mu_1 \geq \mu_2 \geq \cdots \geq \mu_n$, $\nu_1 \geq \nu_2 \geq \cdots \geq \nu_n$, and $\xi_1 \geq \xi_2 \geq \cdots \geq \xi_n$ denote the eigenvalues of $A_n$, $B_n$, and $X_n$, respectively. Then, for all $i, j, k, r, s \in \{1, \dots, n\}$ such that $r + s - 1 \leq i \leq j + k - n$,
\begin{equation*}
\nu_j + \xi_k \leq \mu_i \leq \nu_r + \xi_s.
\end{equation*}
\end{lemma}

\begin{lemma}
\label{lemma:compact}
A compactly supported probability measure on $\mathbb{R}$ is uniquely determined by its moments.
\end{lemma}
\noindent \textit{Proof}. This can be easily proved using \cite{b:bil-08},Theorem 30.1. \hfill $\Box$


\begin{lemma} \label{lemma:maxineq}
Let $A$ and $B$ be $n \times n$ Hermitian matrices. Then,
\begin{equation*}
\max_{1\leq i \leq n} |\lambda_i (A) - \lambda_i(B)| \leq ||A-B||_2
\end{equation*} where $||A||_2 = \sqrt{\lambda_1 (AA^{*})}$ is the spectral norm.
\end{lemma}
\noindent \textit{Proof}. 
By using the Courant-Fischer min-max theorem, for any $n \times n$ Hermitian matrix $A$, we have:
\begin{equation*}
\lambda_i(A) = \sup_{\dim(V) = i} \inf_{v \in V, \|v\| = 1} v^{*} A v, \quad i \in \{1, \dots, n\}.
\end{equation*}
Let $V$ be a subspace with $\dim(V) = i$. For any $v \in V$ with $\|v\| = 1$, we have
\begin{equation*}
v^{*} (A + B) v = v^{*} A v + v^{*} B v \leq v^{*} A v + \|B\|_2.
\end{equation*}
Taking the supremum over all subspaces $V$ with $\dim(V) = i$ and the infimum over all unit vectors $v \in V$, we get
$\lambda_i(A + B) \leq \lambda_i(A) + \|B\|_2$. Therefore, by substituting $A \leftarrow A + B$ and $B \leftarrow -B$ into the inequality, we get
$ \lambda_i(A) \leq \lambda_i(A + B) + \|B\|_2$. This indicates that, for all $i \in \{1, \dots, n\}$, 
$|\lambda_i(A + B) - \lambda_i(A)| \leq \|B\|_2$.
Finally, taking the maximum over $i$ and substituting $B \leftarrow B - A$, we get the desired result.
 \hfill $\Box$